\documentclass{aa}  
\usepackage[french,english]{babel}
\usepackage{amsmath}
\usepackage{graphicx}
\usepackage{graphics}
\usepackage{subfigure}
\usepackage{color}
\usepackage{txfonts}
\usepackage{natbib}
\bibpunct{(}{)}{;}{a}{}{,}

\def\teff{\ifmmode T_{\rm eff} \else $T_{\mathrm{eff}}$\fi}

\def\ltsima{$\buildrel<\over\sim$}
\def\lsim{\lower.5ex\hbox{\ltsima}}
\newcommand{\hi}{H{\sc i}}
\newcommand{\hii}{H{\sc ii}}
\newcommand{\ha}{\ifmmode {\rm H}\alpha \else H$\alpha$\fi}
\newcommand{\hb}{\ifmmode {\rm H}\beta \else H$\beta$\fi}
\newcommand{\lya}{\ifmmode {\rm Ly}\alpha \else Ly$\alpha$\fi}

\newcommand{\McLya}{{\tt McLya}}

\DeclareTextSymbol{\degre}{T1}{6}
\DeclareTextSymbol{\degre}{OT1}{23}

\def\kms{km s$^{-1}$}

\def\msun{\ifmmode M_{\odot} \else M$_{\odot}$\fi}
\def\msunyr{\ifmmode M_{\odot} {\rm yr}^{-1} \else M$_{\odot}$ yr$^{-1}$\fi}
\def\zsun{\ifmmode Z_{\odot} \else Z$_{\odot}$\fi}
\def\lsun{\ifmmode L_{\odot} \else L$_{\odot}$\fi}

\def\mup{\ifmmode M_{\rm up} \else M$_{\rm up}$\fi}
\def\mlow{\ifmmode M_{\rm low} \else M$_{\rm low}$\fi}

\def\aap{A\&A}

\def\apjl{ApJL}
\def\apjs{ApJS}

\newcommand{\oh}{\ifmmode 12 + \log({\rm O/H}) \else$12 + \log({\rm O/H})$\fi}
\def\flyf{\ifmmode f_{\rm Lyf} \else $f_{\rm Lyf}$\fi}
\def\pz{\ifmmode P(z) \else $P(z)$\fi}
\def\ki2{\ifmmode \chi^2 \else $\chi^2$\fi}
\def\zphot{\ifmmode z_{\rm phot} \else $z_{\rm phot}$\fi}

\newcommand{\xphot}{\ifmmode x_\gamma \else $v_\gamma$\fi}
\newcommand{\xobs}{\ifmmode x_{\rm obs} \else $x_{\rm obs}$\fi}
\newcommand{\xcmf}{\ifmmode x_{\rm CMF} \else $x_{\rm CMF}$\fi}
\newcommand{\vexp}{\ifmmode v_{\rm exp} \else $v_{\rm exp}$\fi}
\newcommand{\vmax}{\ifmmode v_{\rm max} \else $v_{\rm max}$\fi}
\newcommand{\nh}{\ifmmode N_{\rm H} \else $N_{\rm H}$\fi}

\begin{document}

\title{Lyman-$\alpha$ emission properties of simulated galaxies: interstellar medium structure and inclination effects}

\author{Anne Verhamme\inst{1,2}, Yohan Dubois\inst{1,3},  Jeremy Blaizot\inst{2}, Thibault Garel\inst{2,4}, Roland Bacon\inst{2}, Julien Devriendt\inst{1,2}, Bruno Guiderdoni\inst{2}, Adrianne Slyz\inst{1}
}
\offprints{anne.verhamme@univ-lyon1.fr}
\institute{
Oxford Astrophysics,
University of Oxford,
Denys Wilkinson Building,
Keble Road, Oxford, OX1 3RH, UK.
\and
Universit\'e de Lyon, Lyon, F-69003, France ;\\ 
Universit\'e Lyon 1, Observatoire de Lyon, 
9 avenue Charles Andr\'e, Saint-Genis Laval, F-69230, France ; \\
CNRS, UMR 5574, Centre de Recherche Astrophysique de Lyon ; \\
Ecole Normale Sup\'erieure de Lyon, Lyon, F-69007, France.
\and
Institut d'Astrophysique de Paris,
98 bis boulevard Arago, 
75014 Paris, France.
\and
Centre for Astrophysics \& Supercomputing, 
Swinburne University of Technology, 
P.O. Box 218, Hawthorn, VIC 3122, Australia.
}
\date{Received date / Accepted date}
\authorrunning{Verhamme et al.}{ }
\titlerunning{\lya\ RT in hydro sims}{ }

\abstract{}
{This paper is the first of a series investigating Lyman-alpha
 (hereafter \lya) radiation transfer through hydrodynamical
 simulations  of galaxy formation. 
 Its aim is to assess the impact of the interstellar medium (ISM) physics 
 on \lya\ radiation transfer and to quantify how galaxy orientation  
 with respect to the line of sight alters observational signatures.
} 
{ We compare the results of \lya\ radiation transfer calculations
  through the ISM of a couple of idealized galaxy simulations in a
  dark matter halo of $\sim 10^{10}M_\odot$. In the first one, G1,
  this ISM is modeled using physics typical of large scale
  cosmological hydrodynamics simulations of galaxy formation, where
  gas is prevented from radiatively cooling below 10$^4$K.  In the
  second one, G2, gas is allowed to radiate away more of its internal
  energy via metal lines and consequently fragments into dense
  star-forming clouds.  }
{First, as expected, {\it the
    small-scale structuration of the ISM plays a determinant role in
    shaping a galaxy's \lya{} properties}. The artificially warm, and
  hence smooth, ISM of G1 yields an escape fraction of $\sim50$\% at
  the \lya{} line center, and produces symmetrical double-peak
  profiles. On the contrary, in G2, most young stars are embedded in
  thick star-forming clouds, and the result is a~$\sim10$ times lower
  escape fraction. G2 also displays a stronger outflowing velocity
  field, which favors the escape of red-shifted photons, resulting in
  an asymmetric \lya{} line. Second, {\it the \lya{} properties of G2
  strongly depend on the inclination at which it is observed}: From
edge-on to face-on, the line goes from a double-peak profile with an
equivalent width of $\sim-5$\AA{} to a 15 times more luminous
red-shifted asymmetric line with EW $\sim 90$\AA{}.}
{The remarkable discrepancy in the \lya\ properties we derived for two
  ISM models raises a fundamental issue. In effect, it demonstrates
  that \lya\ radiation transfer calculations can only lead to
  realistic properties in simulations where galaxies are resolved into
  giant molecular clouds. Such a stringent requirement translates into
  severe constraints both in terms of ISM physics modeling and
  numerical resolution, putting these calculations out of reach of
  current large scale cosmological simulations.  Finally, we find
  inclination effects to be much stronger for \lya{} photons than for
  continuum radiation. This could potentially introduce severe biases
  in the selection function of narrow-band \lya{} emitter surveys, and
  in their interpretation, and we predict these surveys could indeed
  miss a significant fraction of the high-$z$ galaxy population.
} 

\maketitle

\section{Introduction}
\label{s_intro}

In the last decade, the Lyman-alpha (\lya{}) emission line has become
an observational tool of choice to detect high redshift galaxies via
narrow-band surveys
\citep[e.g.][]{Hu1998,Kudritzki2000,Shimasaku2006,Ouchi2008,Ouchi2010,Hu2010}
or blind spectroscopic searches
\citep[e.g.][]{vanBreukelen2005,Rauch2008,Cassata2011}. Today, the number of
galaxies detected in this fashion (hereafter Lyman-Alpha Emitters, LAE)
is becoming statistically significant, and LAEs play a major role in
our census of high-$z$ galaxies. At the same time, spectroscopic
follow-ups of UV-selected galaxies shed more and more light on the
physical nature of LAE and on their place in the cosmic history of
galaxy formation \citep{Shapley03,Tapken07,Bielby11}. 
One of
the major challenges in the years to come, both theoretical and
observational, is yet to understand the details of the \lya{} line
profiles we observe: How do they relate (if they do) to any physical
property of high-$z$ galaxies~?  

Although a number of semi-analytic models for Lyman-alpha Emitting galaxies (LAEs)
have been published
\citep[e.g.][]{Ledelliou06,Orsi08,Dayal2008,Dayal2009,Orsi2011,Dayal2011,
  Garel2012}, the complete radiation transfer through 
\emph{the interstellar medium} of \lya\ emitting galaxies 
has been taken into account in only a handful of previous studies 
\citep[][see Table \ref{compar} for a summary.]
{tasitsiomi06,Laursen2009, Barnes2012,Yajima2011,Yajima2012}. 

\begin{table*}
\begin{tabular}{c|c|c|c|c|c}
               & this study              & \citet{tasitsiomi06}    & \citet{Laursen2009}     & \citet{Barnes2012}     & \citet{Yajima2011,Yajima2012}       \\
\hline
context        & \lya\ emitting galaxies & \lya\ emitting galaxies & \lya\ emitting galaxies & DLA-host galaxies      & \lya\ emitting galaxies  \\
\hline
hydro technics & AMR (RAMSES)            & AMR (ART)               & SPH (TreeSPH)            & SPH (GADGET)           & SPH (GADGET)             \\
\lya\ RT       & \lya\ + continuum       & \lya, no dust           & \lya                    & \lya, no dust & \lya\ + continuum        \\
               &    AMR                  &   AMR                   & AMR                     & cartesian              & AMR                      \\
\lya\ sources  & recombination           & recombination           & recombination           & central point source   & recombination            \\
               & from young stars        & from young stars        & + gravitational cooling &                        & + collisional excitation \\
               &                         &                         & + UV background         &                        &                          \\
environment    & isolated galaxy         & cosmo zoom              & cosmo zoom              & cosmo zoom             & cosmo zoom               \\
nb of objects  & 2 & 1 & 9 & 3 & $\sim 950$ \\
stellar mass   & $1.8\times10^9$\msun for G1 & $\sim 10^{10}$\msun & $6\times10^6$ to         & $1.5\times10^{10}$\msun & $4.3\times10^9$\msun     \\
               & $4.9\times10^8$\msun for G2 &                     & $3\times10^{10}$\msun    & $1.5\times10^{11}$\msun & $9.3\times10^9$\msun     \\
               &                             &                     &                         & $7.5\times10^{11}$\msun & $4.1\times10^{10}$\msun   \\
stellar mass resolution & $1.4\times10^3$\msun for G2 & $2\times10^4$\msun & $10^6$\msun           & not available          & $1.9\times10^4$\msun      \\
                        & $7.7\times10^3$\msun for G1 &                    &             &               & \\
spatial resolution$^{(a)}$& 18 pc for G2            & 29 pc               &
137 pc$^{(b)}$                  & 514 pc              & 342 pc                     \\ 
                   & 147 pc for G1           &                     &                         &                        &                            \\
\hi\ temperature   & $10^2$ to $10^5$K       & $10^3$ to $10^4$K    & $10^4$ K                & $10^{4.3}$ to $10^5$K   & -              \\
\end{tabular}
\caption{Comparison of the 4 published studies of \lya\ radiation transfer through 
  the interstellar medium of galaxies with this study. \newline
  $^{(a)}$ Resolution, in physical $pc$. This is either the minimum cell size, for
  AMR codes, or the gas gravitational softening length for SPH codes. In
  both cases, this reflects the smallest scale onto which a gas overdensity
  may feel its own gravity. \newline
  $^{(b)}$ This resolution corresponds to their S87 simulation, which
  best matches our halo mass.} 
\label{compar}
\end{table*}

In \citet{tasitsiomi06}, \lya\ radiation transfer is post-processed in the
 brightest \lya\ emitter of a gas-dynamics+N-body adaptive refinement
 tree \citep[ART,][]{Kravtsov05} simulation at $ z\sim 8$, in order to investigate its 
detectability.
In \citet{Laursen2009}, a sample of nine galaxies at $z = 3.6$ taken from  
cosmological N-body/hydrodynamical TreeSPH simulations \citep{Sommer06} 
are considered, sampled in mass. 
The main result of the paper is an anti-correlation between \lya\ escape 
fraction and the mass of the galaxy.
In \citet{Barnes2012}, three halos at $z = 3$ are selected in cosmological
hydrodynamic simulations (GADGET-2) aimed at reproducing the physical properties 
of the host galaxies of DLAs at $z\sim 3$ \citep{Tescari09}.
In \citet{Yajima2011,Yajima2012}, the same halo is followed at different redshifts, 
in order to study the evolution of the \lya\ properties of 
their galaxies with time.

These previous studies are done through a warm interstellar medium in which 
the gas is cooled down to T$\sim10^4$K, with the consequence that the formation 
of small scale structures is not modelled.
Indeed, the pressure support of this warm gas prevents it from collapsing at 
scales smaller than its Jeans length, explaining the difference in spatial resolution 
between the different experiments (see section \ref{s_hydro} for more details).
However, theoretical expectations suggest a strong dependance of the \lya\ 
transfer on the structure, and geometry of the interstellar medium of 
galaxies, and the main goal of our study is to investigate this point.  
Furthermore, two studies are dust-free, which prevent them from studying the \lya\ escape 
fraction from their configurations.
Finally, monochromatic approaches of the problem do not allow to derive
\lya\ equivalent widths, and compare the transfer of continuum versus line
photons.

To overcome these limitations, we post-process hydrodynamical simulations of
galaxy formation described in \citet{Dubois08} performed  with the RAMSES code 
\citep{Teyssier02}, with \McLya{} \citep{verhamme06}, including \lya+continuum 
radiation transfer in a dusty medium, for two different ISM models :
1/ the reference model G1, comparable to previous studies, where the gas is cooled down to $10^4$K,
2/ a more realistic ISM model G2, where the gas is allowed to cool down to 100K, and the formation 
of small scale structures is followed.  

The plan of this paper is the following. 
We start to describe the hydrodynamical simulations 
used to post-process \lya\ radiation transfer. Then, we 
describe the radiative transfer of \lya\ photons in the hydrodynamical simulations 
with \McLya. The fourth part presents a comparison of the \lya\ properties 
for the two ISM models.
In the fifth part, we discuss the effect of orientation on the \lya\ properties of 
G2, which appears as an interesting bonus result of this work.
The sixth part describes the diffuse \lya\ halo around G2.
The last part summarizes the main conclusions.

\section{Description of the hydrodynamical simulations}
\label{s_hydro}

The results presented in this paper are based on the analysis of a  
couple of
idealized high-resolution hydrodynamical simulations which follow the
formation and evolution of an isolated disc galaxy embedded in a live  
dark
matter (DM) halo. Both simulations were run
with RAMSES \citep{Teyssier02}, using sub-grid physics modules  
described in
\citet{rasera&teyssier06} and \citet{Dubois08}.

\subsection{Initial Conditions} \label{sec:ICs}

We choose to follow the formation of a galaxy embedded
in a halo of total mass (dark matter plus gas) $M_{200}=10^{10}\, \rm
M_\odot$. The choice of this rather small halo mass is motivated
by the expectation that most small-mass haloes host LAEs, while only a
fraction of massive galaxies emit Lyman-alpha at all \citep[see
e.g.][and references therein]{Garel2012}.

We assume a NFW density profile \citep{navarroetal96} with  
concentration parameter
$c=10$ for the DM halo. The `virial' radius $R_{200}\sim 35$ kpc of  
this latter is defined so that
it encloses an average density equal to 200 times the mean matter  
density of
the Universe at $z=0$, assuming $\Omega_m=1$ and $h=1$. Note that with
the currently favored cosmology \citep[e.g.][]{WMAP7}, this value of
$R_{200}$ corresponds to a $z\sim 1$ halo of the same mass.

We follow \citet{Dubois08} to generate our idealized initial
conditions and sample the DM halo with particles instead of
using a static gravitational potential. This is necessary to allow gas  
and stars
to exchange angular momentum with the DM. This process is particularly
important in simulations where gas fragmentation occurs in the disc
because the resulting star forming and gravitationally bound clouds  
should be driven
to the bottom of the potential well by dynamical friction. In  
practice, we sample our halo out to a radius
of $3.2\times R_{200}$ with $10^5$ DM particles of mass $\sim 1.8\
10^5 \rm M_\odot$.

In order to form a centrifugally supported gas disc, we give this  
distribution of DM
particles a specific angular momentum profile $j(r) = j_{\rm
  max}\times M(<r)/M_{200}$, where $r$ is the distance to the halo
center. The normalization $j_{\rm max}$ is fixed by setting the
dimensionless spin parameter of the halo to $\lambda = 0.04$ \citep[see
e.g.][]{bullocketal01}.
The initial gas density profile is obtained by scaling that of the DM by
the universal baryonic fraction $f_{\rm b}=0.15$. The gas velocity  
field is chosen to be
identical to that of the DM. We enforce
hydrostatic equilibrium, which uniquely defines the gas initial  
temperature (or
pressure) profile, and its chemical composition is pristine (76 \%  
Hydrogen and 24 \% Helium in mass).

\vskip 0.2cm

The simulated volume is a cubic box of size $L_{\rm box}=300$ kpc on a  
side, sufficiently large to self-consistently follow
the collapse of the gaseous halo for up to 6 Gyrs. Open boundary  
conditions let the galactic wind stream out of the box freely.
The snapshots are chosen at the time where the 
galactic wind is reaching the virial radius, i.e. after 3 Gyr for G1 and 1 Gyr for G2.
Although ages are different, M$_{\rm gas}$ and M$_{\rm dust}$ are
comparable for G1 and G2 (See Table~\ref{t_G1G2}).
The adaptive mesh refinement (AMR) grid is composed of $64^3$ cells on  
the coarsest level, which corresponds to a
cell size of $\sim4.7$ kpc. Levels of refinement are triggered when at  
least one of the two following criteria is fulfilled:
\begin{enumerate}
\item The total baryonic mass in a cell becomes larger than $6\, 10^4  
\, \rm M_{\odot}$, or a cell contains at least 8 DM particles
\item A cell size becomes larger than $0.25 \lambda_{\rm J}$  
\citep{truelove}
\end{enumerate}
Refinement is allowed up to a maximum level which depends on the  
simulation. In our G1 (resp. G2) simulation
(see Tab. \ref{compar} and Sec.  \ref{sec:physics}), the maximum level  
allowed is 11 (resp. 14), which corresponds to a
spatial resolution of 147 pc (resp. 18 pc).

\begin{figure*}
\begin{tabular}{ccc}
\includegraphics[width=8cm]{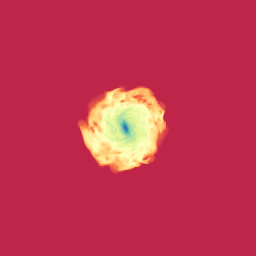}
\includegraphics[width=8cm]{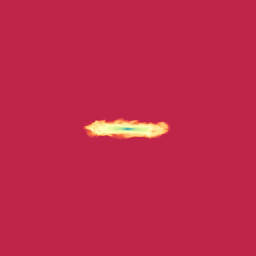}
\includegraphics[width=1.5cm]{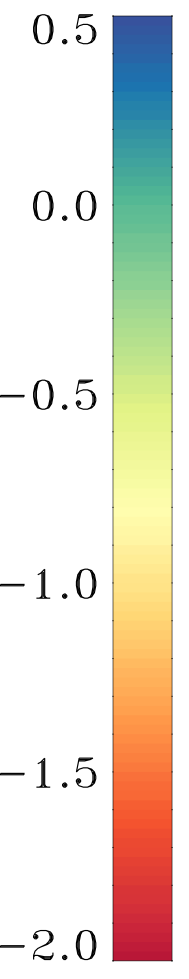} \\
\includegraphics[width=8cm]{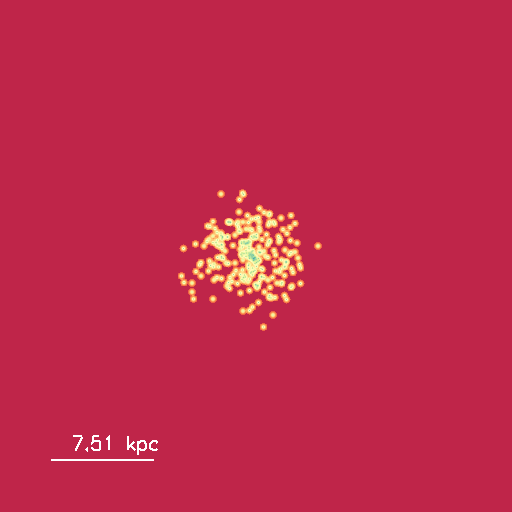}
\includegraphics[width=8cm]{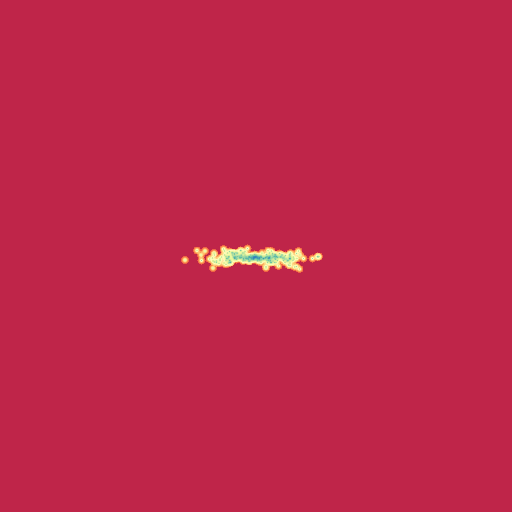}
\includegraphics[width=1.5cm]{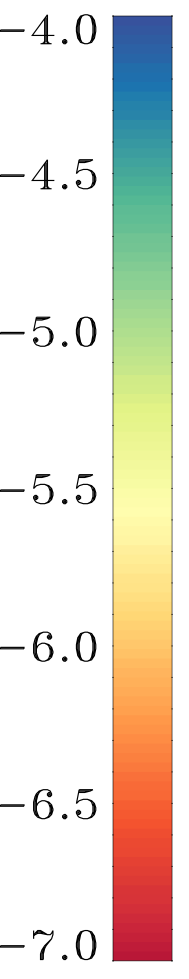}
\end{tabular}
\caption{{\bf Left :} G1 run galaxy seen face-on. {\bf Right :} G1 run  
galaxy seen edge-on.
The upper panels show the distribution of the gas density (in log  
$cm^{-3}$ units), and the bottom panels the
distribution of stellar density for stars younger than 10 Myr (in  
arbitrary log units) .}
\label{wind}
\end{figure*}

\begin{figure*}
\begin{tabular}{ccc}
\includegraphics[width=8cm]{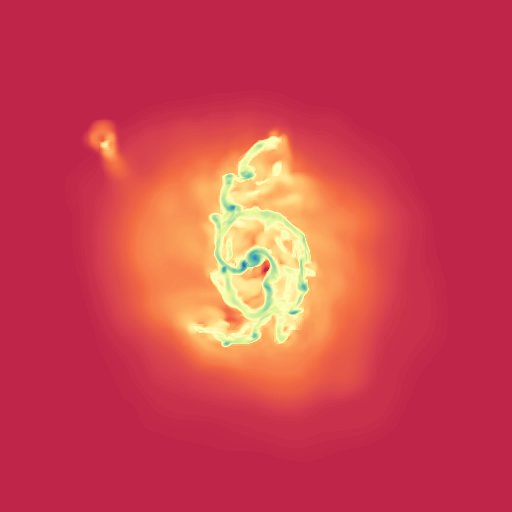}
\includegraphics[width=8cm]{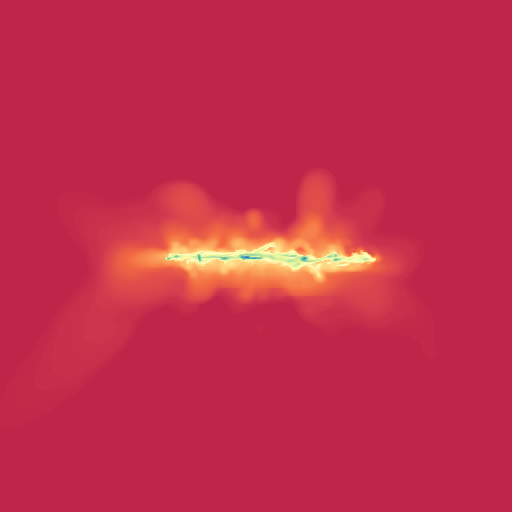}
\includegraphics[width=1.5cm]{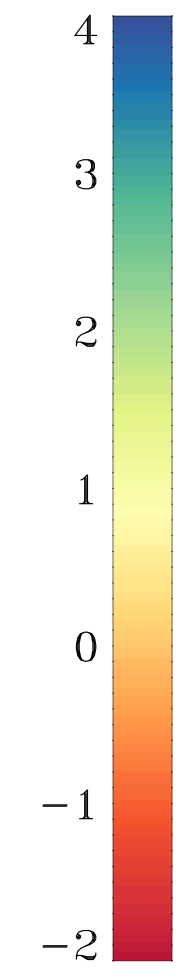} \\
\includegraphics[width=8cm]{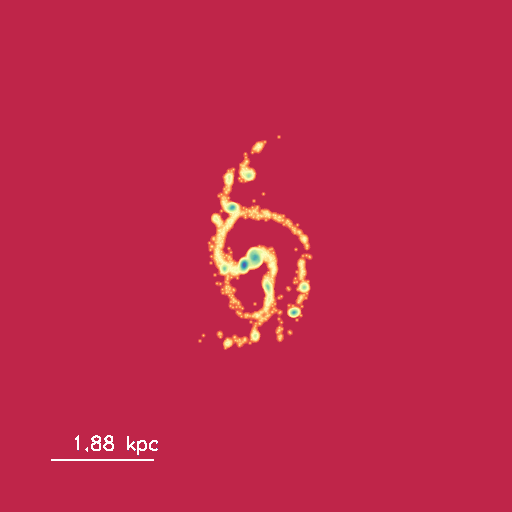}
\includegraphics[width=8cm]{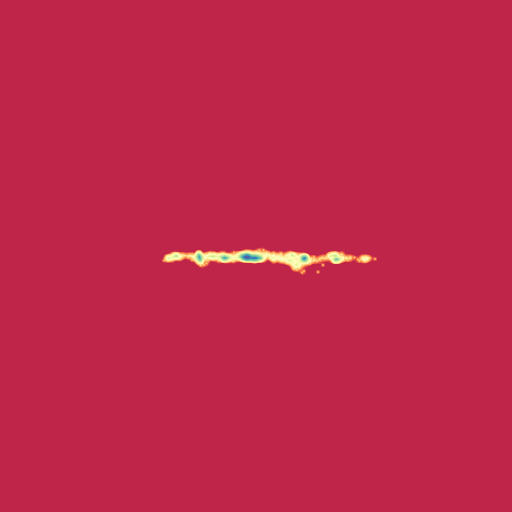}
\includegraphics[width=1.5cm]{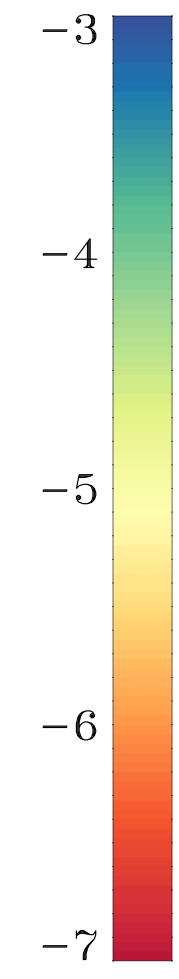}
\end{tabular}
\caption{Same as Fig.~\ref{wind} for the G2 run.}
\label{clumps}
\end{figure*}

\subsection{Physics of galaxy formation} \label{sec:physics}

In this subsection, we briefly discuss how the physics (besides  
gravity and hydrodynamics) relevant to galaxy formation
is modeled in the simulations.
Table~\ref{t_G1G2} summarizes the global properties of G1 and G2.

\begin{table}
\begin{tabular}{c|c|c}
                       &        G1            &        G2           \\
\hline
Stellar mass M$\star$  & $2.1\times10^9$\msun & $6.8\times10^8$\msun \\
Gas mass M$_{\rm gas}$  & $5.0\times10^8$\msun & $3.0\times10^8$\msun \\
M$\star$/M$_{\rm gas}$  &      $\sim4$         &         $\sim2$      \\
Dust mass M$_{\rm dust}$& $8.7\times10^6$\msun & $6.3\times10^6$\msun  \\
SFR                    &       0.3\msun/yr    &     1.1\msun/yr      \\
\end{tabular}
\caption{Comparison of the physical properties of G1 and G2. 
From the SFR, and assuming a Salpeter IMF,
 we can derive intrinsic \lya\ luminosities of our galaxies:
$L(\lya)_{\rm G1} = 3.3\times10^{41} erg.s^{-1}$ and 
$L(\lya)_{\rm G2} = 1.2\times10^{42} erg.s^{-1}$.}
\label{t_G1G2}
\end{table}

\noindent {\bf Cooling:} Radiative energy losses, assuming gas is in  
collisional ionization equilibrium, are computed in each grid cell   
using the metallicity dependent cooling functions tabulated in  
\citet{sutherland&dopita93}. This allows the gas to cool down to  
10$^4$K in runs G1 and G2. In G2, we also account for the extra loss  
of energy provided by metal line cooling, following the prescription  
of \citet{rosen&bregman95}. This allows the temperature of the gas to  
drop further down, to 100K in this run.

\vskip 0.2cm
\noindent {\bf Interstellar Medium:} In order to prevent numerical  
fragmentation, we assume the ideal gas transitions to a polytropic  
equation of state (EoS) in high-density regions
\begin{equation} \label{eq:polytrope}
T=T_0 \left({\rho \over \rho_0}\right)^{p-1}\, ,
\end{equation}
where $p$ is the polytropic index, $\rho_0$ is the gas density above  
which the polytropic law applies, and $T_0$ is the minimum temperature  
of the gas due to cooling. We adopt a $p=2$ polytropic index that  
ensures that the Jeans length is independent of the gas density for  
densities above $\rho_0$. For  
run G1,  $T_0=10^4$ K and $\rho_0=0.1\, \rm H.cm^{-3}$, and the Jeans  
length of the gas is $\lambda_{\rm J}\simeq 5.8$~kpc, comparable to  
the size of the entire galactic disc.  This means that the ISM gas  
does not fragment, but instead is smoothly distributed throughout the  
disc (see Fig~\ref{wind}).
For run G2,
$T_0=100$ K, and $\rho_0=10\, \rm H.cm^{-3}$, so the constant Jeans  
length given by the polytropic EoS is $\lambda_{\rm J}=58$~pc, which  
we resolve with $\sim 3$ cells of size
 $\Delta x=18$ pc.  This choice of $T_0$ and $\rho_0$ creates a  
multiphase medium with cold and dense bounded regions of size $\sim  
100$ pc similar to the large
 giant molecular clouds (GMCs) of the Milky Way, and a low-density  
warm medium with strong turbulent motions (see Fig~\ref{clumps}).  

\vskip 0.2cm
\noindent {\bf Star formation:} Star formation is modeled as in
\citet{rasera&teyssier06}, with a random Poisson
process spawning star cluster particles according to a classic Schmidt  
law.
In other words, the star formation rate $\dot \rho_*$ scales with the  
local gas
density $\rho$ as
\begin{equation}
\dot \rho_*=\epsilon_* {\rho\over t_{\rm ff}} \, ,
\end{equation}
where $t_{\rm ff}$ is the free fall time and $\epsilon_*$ is the star  
formation efficiency. We adopt $\epsilon_*=1\%$ in runs G1 and G2,  
which is in fair agreement with
observations over the gas density range we span   
\citep{krumholz&tan07}.  Star formation is only permitted to occur in  
grid cells where the gas density is larger than $\rho_0$.
Each star-cluster particle created is given a mass which is an integer  
multiple of the minimal mass $m_*=\rho_0\times \Delta x^3$, where $ 
\Delta x$ is the size of a cell on
the highest level of refinement. Therefore, our stellar mass  
resolution is $m_*\simeq1.4\, 10^3\, \rm M_{\odot}$ for the G2  
simulation (and $m_*\simeq7.7\, 10^3\, \rm M_{\odot}$ for the G1  
simulation).  We forbid the star formation process to consume more  
than $90\%$ of the gas mass in the cell where it takes place during a  
time step.

\vskip 0.2cm
\noindent {\bf Feedback:} We model supernovae (SN) feedback as in  
\cite{Dubois08}, i.e. we deposit kinetic energy in a sphere centered  
on the explosion, along with mass and momentum distributed according  
to a Sedov-Taylor blast-wave profile.  This kinetic approach ensures  
the formation of large-scale galactic winds in low-mass haloes, and  
galactic fountains in the most massive ones even at low resolution  
(\citealp{Dubois08}). Such a precaution is necessary since it is well  
known that spurious thermal energy losses are catastrophic  
\citep{navarro&white93}.  We assume a standard \cite{salpeter55}  
Initial Mass Function (IMF), with $\eta_{\rm SN}=10\%$ (in mass) of  
the newly formed stars finishing their $\sim$ 10 Myr lives as type II  
SN.  We further assume that $10^{51}$ ergs of energy are released by  
each individual SN explosion with a typical ejected mass of 10  
$M_{\odot}$.  SN are also responsible for releasing metals into their  
surroundings with a constant yield $y=0.1$.  These metals are  
passively advected.

\section{\lya\ transfer}
\label{s_mclya}

We use an improved version of the Monte Carlo radiation transfer code  
\McLya{} of \citet{verhamme06} to post-process the two simulations G1  
and G2 described in Sec. \ref{s_hydro}. The most notable technical  
improvement over the original \McLya{} is that it now fully exploits  
the AMR grid structure of RAMSES, which allows us to perform the  
radiative transfer at the same spatial resolution than our  
hydrodynamics simulations. The new version of \McLya{} also features  
more detailed physics of the \lya{} line and UV continuum transfer,  
which include \citep[see][]{Schaerer2011}: angular redistribution  
functions taking quantum mechanical effects for \lya\ into account  
\citep{dijkstra08,stenflo80}, frequency changes of \lya\ photons due  
to the recoil effect \citep[e.g.][]{zheng02}, the presence of  
deuterium \citep[assuming a canonical abundance of $D/H = 3 \times  
10^{-5}$,][]{Dijkstra06}, and anisotropic dust scattering using the  
Henyey-Greenstein phase function \citep[using parameters adopted in][] 
{witt00}.

Before \McLya{} can be used to process our simulations, we have to  
extract a set of gas properties for each cell which are not directly  
predicted by RAMSES: the gas velocity dispersion due to temperature  
and small-scale turbulence (the macroscopic velocity and temperature  
of each cell are known), the ionization state (or equivalently the  
neutral Hydrogen density), and the dust content. We discuss how we  
compute these quantities in the following sub-sections, and then  
explain our strategy to sample \lya{} emission.

\subsection{Ionisation state of the gas} \label{sec:ionization}
The simulation outputs provide us with the total density $\rho$ of gas in each cell, including H, He  
and metals. We derive the numerical density of H as $n_H = X_H\ \rho / m_H$, where $X_H=0.76$ is the mass fraction of H, and $m_H$ the mass of an H atom. \lya{} photons however only scatter onto  
neutral Hydrogen atoms (\hi), the numerical density of which we compute as 
$n_{\rm HI} = x \ n_H$,  
where the neutral fraction $x$ is evaluated  
assuming collisional ionization equilibrium (CIE) and hence only  
depends on temperature $T$.

Using the polytropic equation of state (Eq. \ref{eq:polytrope}) to  
prevent artificial fragmentation in the dense ISM unfortunately makes  
the gas temperature in these regions unknown. In all polytropic cells,  
we therefore set the temperature of the gas to $T_0$ ($= 10^4$~K in G1  
and $10^2$~K in G2). Since we are assuming CIE, these temperatures  
imply that all the polytropic gas is neutral.

\subsection{Velocity dispersion of the gas}

The transfer of \lya\ photons through a parcel of gas depends on the  
velocity distribution of the neutral hydrogen atoms in that gas. This  
dependence appears in the Voigt profile through the Doppler parameter  
$b$. In the
case of purely thermal motion, $b = v_{\rm th} = (2k_BT/ 
m_H)^{1/2}$. In the presence of small-scale turbulence, one has to  
quadratically add the turbulent velocity
($v_{\rm turb}$) contribution and write $b = (v_{\rm th}^2 + v_{\rm turb}^2)^{1/2}$.

In practice, we set $v_{\rm turb} = 10$ km s$^{-1}$ in both G1 and G2  
simulations. This value is the mean value of turbulent velocities  
computed in simulations of
SN-driven turbulence by \citet{dibetal06}. We tested the  
robustness of our results against this assumption by repeating the  
calculations presented in this paper
with  $v_{\rm turb} = 0$ or $v_{\rm turb} = v_{\rm th}$. We found no  
impact on the \lya{} escape fractions.

\subsection{Dust distribution}

In our simulations, gas metallicity is self-consistently calculated  
from the release of metals in the ISM by SN explosion.  The dust  
distribution is derived from the metallicity distribution by assuming a  
galactic value for the dust-to-metal (mass) ratio R$_{\rm dust/metal}$ = 0.3  
\citep{Inoue03}.  We further assume that dust is present only in neutral (i.e. $T\lsim 10^4$~K) media, and write:
\begin{equation}
n_{\rm dust} = R_{\rm dust/metal} \frac{Z}{X_H} \frac{m_H}{m_d} n_{\rm HI},
\end{equation}
where $Z$ is the metallicity,
$m_H/m_d = 5 \times 10^{-8}$ is the proton to dust particle mass  
ratio \citep{Draine84}, and
$n_{\rm HI}$ is the numerical density of {\it neutral} hydrogen  
derived in Sec. \ref{sec:ionization}.

\subsection{The \lya\ sources}
\label{s_input_spec}

\begin{figure}
\includegraphics[width=8.6cm]{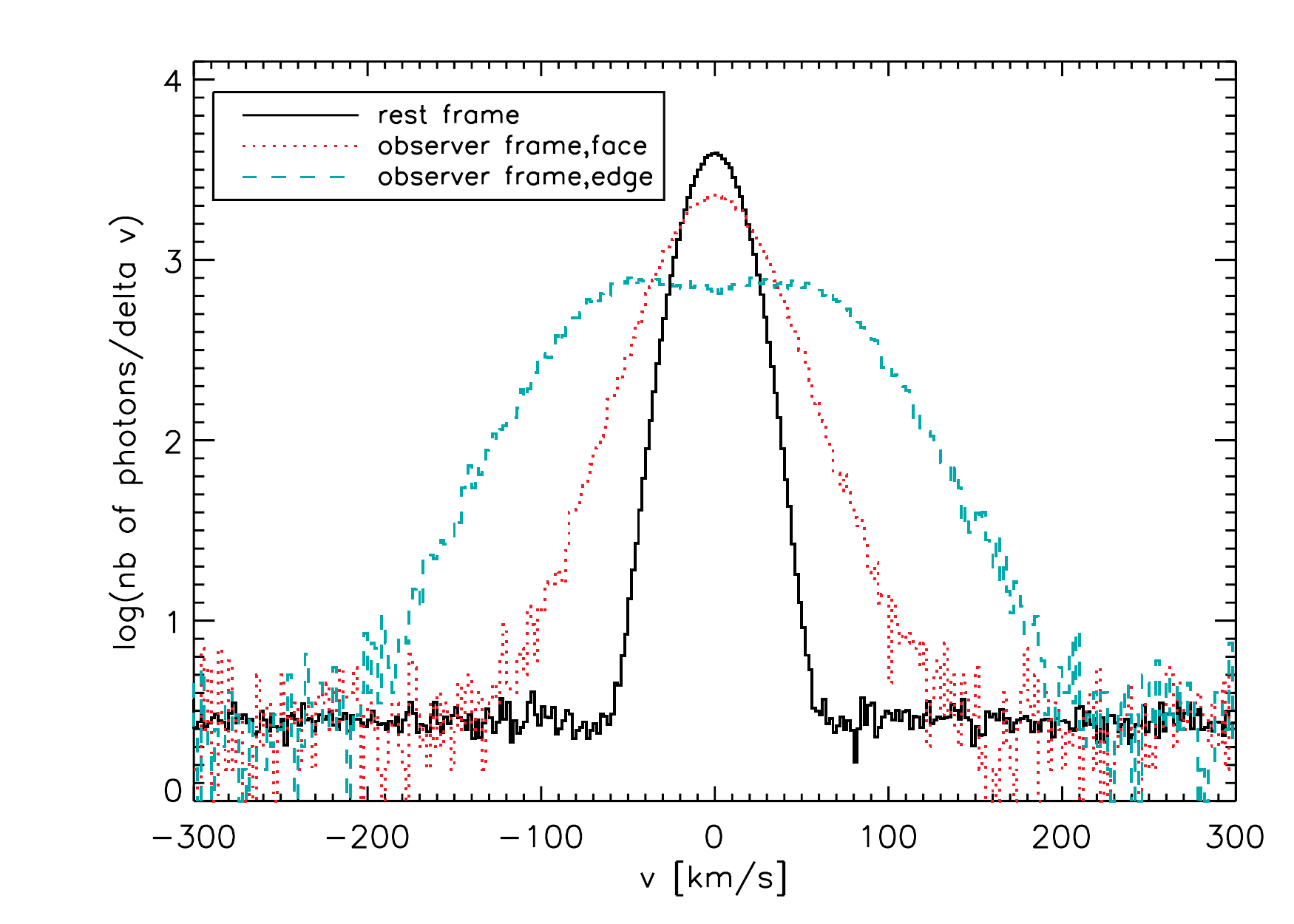}
\caption{Number of photons emitted per unit velocity bin. The black  
histogram shows the frequency distribution in the sources frame, as  
sampled by all photons used in our G2 experiment. This line profile is  
the one we use as a model for \hii{} regions (Sec.  
\ref{s_input_spec}). The dotted red (resp. dashed blue) histograms  
show how these photons are
distributed in the observer frame when the galaxy is seen face-on  
(resp. edge-on).  }
\label{fig:input_spec}
\end{figure}

In the present work, the only mechanism which produces \lya{} photons  
that we take into account is the recombination of photo-ionized gas  
surrounding massive stars \citep{Kennicutt98}.
In practice,
given our resolution, we are effectively using young star particles as  
a proxy for \hii{} regions and shooting photons from their locations.  
Note that we only consider star particles younger than 10 Myr as they  
are the only ones producing significant amounts of ionizing photons  
\citep[e.g.][]{CharlotFall93}.

Each one of these young star particle emits on average $N_{\rm phot} /  
N_{\star}$ photons\footnote{We deliberately elect not to fix the  
number of photons per source, but
the total number of photons for the whole experiment and sample  
sources randomly. Thus, the number of photons emitted by each  
individual star particle has Poissonian variance.}, where $N_{\rm phot} 
$ is the total number of photons in the experiment, and $N_{\star}$ is  
the number of young stellar particles in the simulation. In our G1  
simulation, we fix $N_{\rm phot} = 6.4\ 10^5$, and since this  
simulation has $N_\star \sim 10^3$, this implies an average of 640  
photons per source. In our G2 simulation, we pick $N_{\rm phot} \sim  
5.1\ 10^6$, and since G2 contains $N_\star \sim 3\ 10^4$, this yields  
an average of 170 photons per source.

In the rest-frame of a star particle, the photons are emitted so as to  
sample a profile defined by a flat continuum plus a Gaussian line of  
full-width at half maximum of 20 \kms, and of equivalent width  
EW(\lya) = 200 \AA{} \citep[as suggested by e.g. ][] 
{CharlotFall93,Valls-Gabaud93,Schaerer03}. The continuum extends from  
-20000 \kms{} to 20000 \kms{} around the \lya{} line in the star  
particle rest-frame, which is way beyond any resonant transfer effect.  
As an illustration, we show on Fig. \ref{fig:input_spec} the source  
frame emission profile (black histogram) along with the emitted  
spectrum (emission from all star particles) of the face-on (red  
histogram) and edge-on (blue histogram)
G2 galaxy in the observer frame.

\section{How the ISM structure impacts \lya{} transfer}
\label{s_results}

The different temperature floors of the cooling functions used in G1  
($10^4$ K) and G2 (100 K) result in strikingly different  
structurations of the ISM. While the gas in G2 is able to fragment  
into small star-forming clumps, the thermal pressure support in G1  
yields a rather smooth ISM with homogeneous star formation. In the  
present section, we investigate the influence of these structures on  
the \lya{} escape fractions and line profiles.

\subsection{Escape fractions}

\begin{figure}
\begin{tabular}{l}
\includegraphics[width=8.6cm]{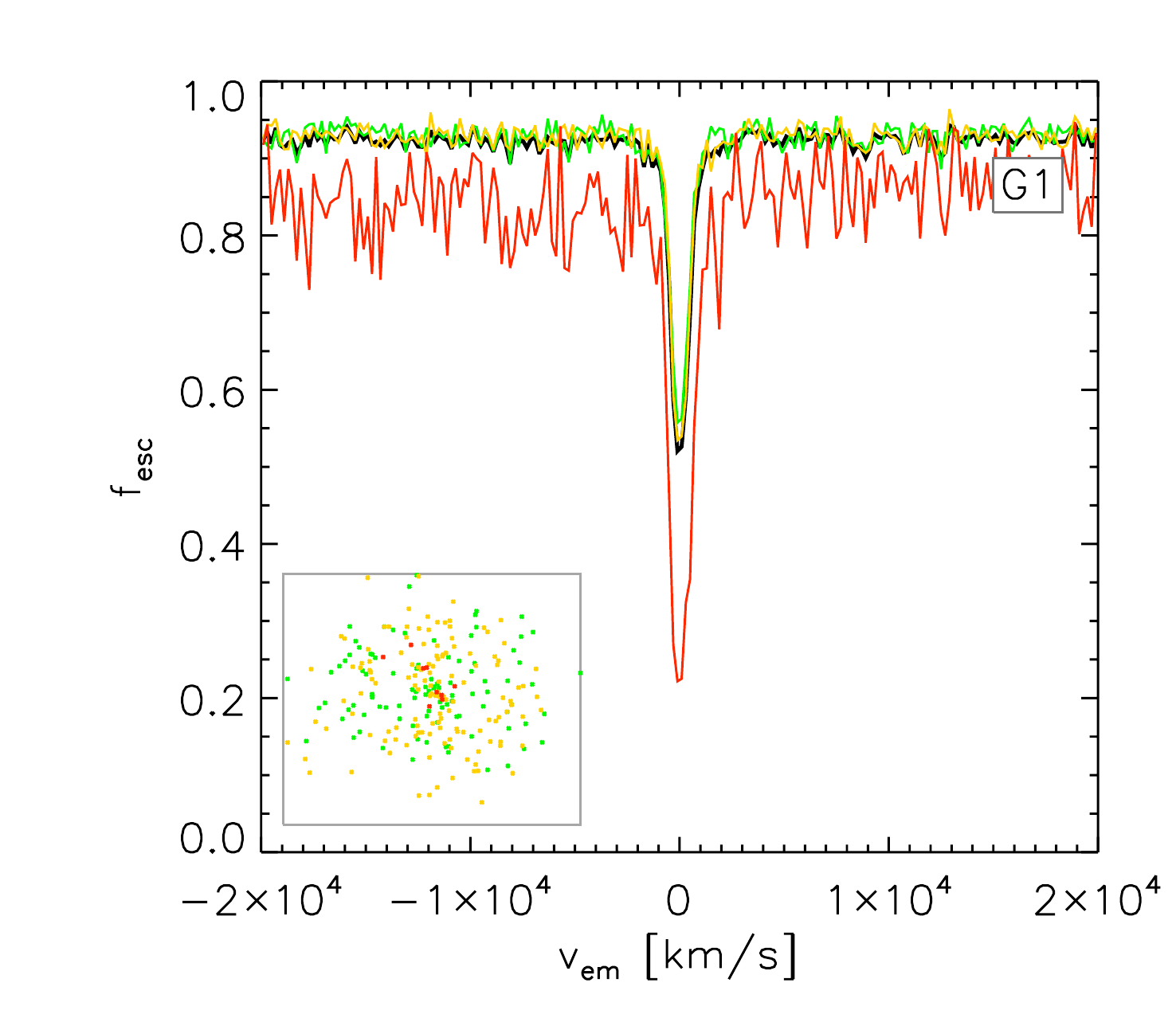}\\
\includegraphics[width=8.6cm]{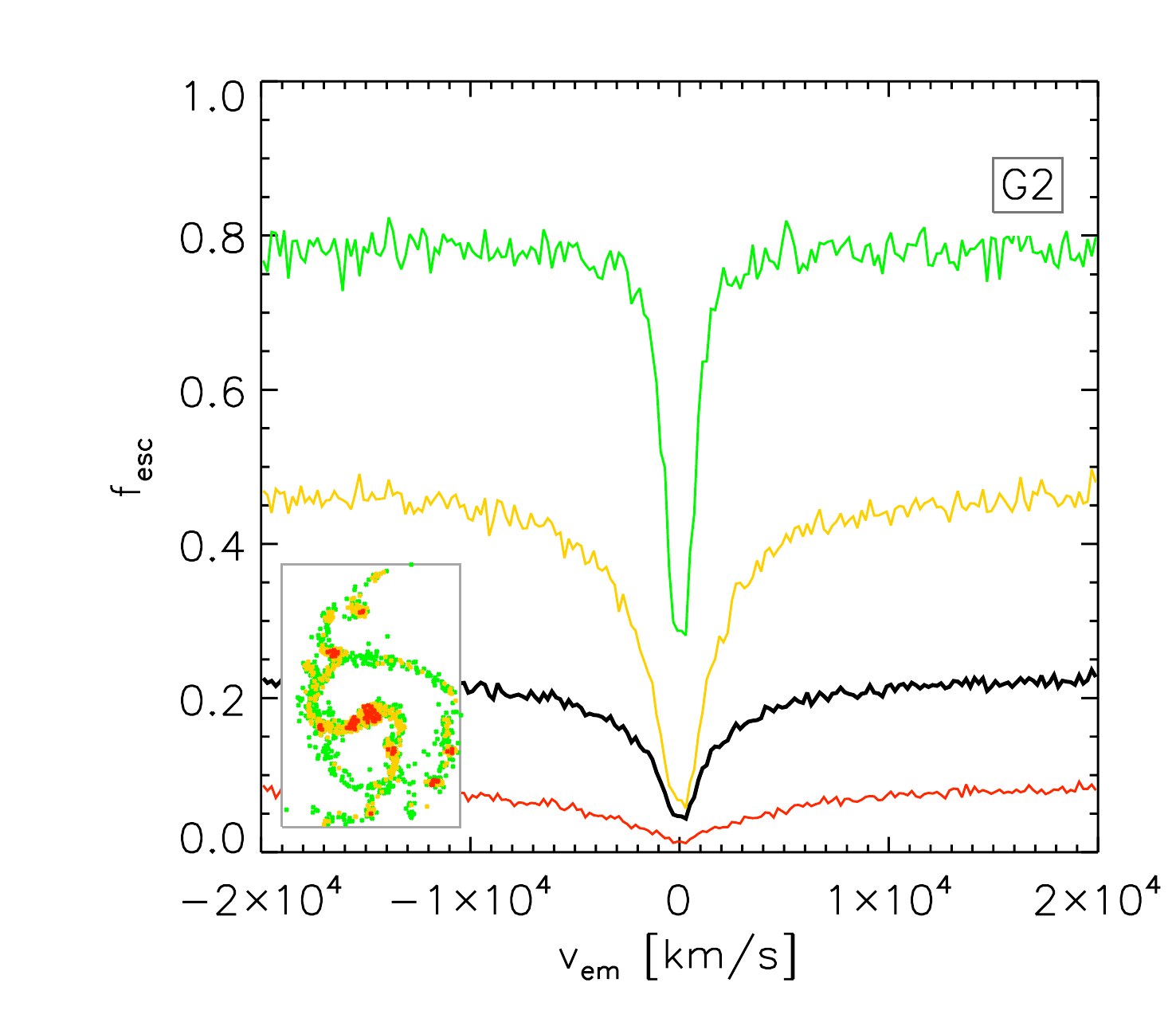}
\end{tabular}
\caption{Escape fraction as a function of emitted frequency $v_{\rm em}$ 
(in the source rest-frame) for simulation G1 (top) and G2 (bottom).  
The black curves show the spatially integrated escape fractions, i.e.  
escape fractions averaged over all sources in the galaxy. The red  
(resp. yellow, green) curves show the escape fractions for sources in  
the high (resp. average, low) density regions. Thumbnails in the lower-left corner of each  
panel indicate the face-on position of sources contributing to each of  
the three curves of the same color.}
\label{fesc_jeje}
\end{figure}

Let us first look at the escape fraction as a function of emission  
frequency $f_{\rm esc}(v_{\rm em})$, defined as the fraction of  
photons emitted at a frequency $v_{\rm em}$ (in the source rest-frame)  
which eventually escape the galaxy (i.e. which are not absorbed by  
dust), whatever their observed frequency. The black curves on Fig.  
\ref{fesc_jeje} show $f_{\rm esc}$ as a function of $v_{\rm em}$ for  
G1 (top panel) and G2 (bottom panel).

Before we discuss these curves in more detail, we can partially  
compare our results to those of \citet{Laursen2009}. The simulations of  
these authors are relatively similar to our G1 simulation in the sense  
that their cooling curve also stops at 10$^4$ K
\citep{Sommer-Larsen03}
and therefore their ISM is unstructured on small scales. As  
\citet{Laursen2009} only propagate \lya{} photons emitted exactly at the  
\lya{} frequency, we have to compare their escape fractions to our  
value at $v_{\rm em} = 0$, which is about 55\%. This is broadly  
consistent with their results (their Fig. 9, at $M_{\rm vir} =  
10^{10}$ M$_\odot$).

We now turn to the comparison of the escape fractions predicted for G1  
and G2 (black curves in Fig. \ref{fesc_jeje}). The first remarkable  
difference is seen in the {\it continuum} escape fractions 
\footnote{Note that the continuum here is only that produced by star  
particles younger than 10 Myr, and not by the full stellar population  
of each galaxy.}, which are $\sim 95$\% in G1 versus $\sim 22$\% in  
G2. This difference has a double origin. The first origin is the very  
different ISM structures found in G1 and G2. As gas in G2 is able to  
cool to lower temperatures, it fragments into large star-forming  
clouds within which most young stellar particles are buried. Instead,  
gas in G1 has a stronger pressure support and star formation is more  
diffuse and young stars end up distributed in a rather low density  
environment. Indeed gas density in G1 does not reach values above 10  
atoms per cm$^3$, while gas in G2 reaches 10$^4$ atoms per cm$^3$. The  
second origin is that dust, which we model to scale with the density  
of metals (and neutral hydrogen) naturally follows star formation.  
Although it mixes on large scales with time, the dense star-forming  
clouds of G2 turn out to be more dust-rich than the diffuse (well  
mixed) ISM of G1. This produces more extinguished young stellar  
populations in G2, in agreement with e.g. \citet{CharlotFall00}.

The second, more subtle point to take form Fig. \ref{fesc_jeje} is  
that the escape fraction of \lya{} photons is also less in G2 than in  
G1, even when normalized to that of the continuum. Indeed, the ratio  
of line center to continuum escape fractions in G1 and G2 are  
respectively $\sim 54$\% and $\sim 20$\%. We interpret this as a  
natural effect of resonant scattering of \lya{} photons. In the dense  
clumps of G2 where most photons are emitted, the effect of dust is  
enhanced by the large \hi{} column densities which increase the path  
of photons and hence their likeliness to hit a dust grain. The \lya{}  
escape fractions at the line center vary by an order of magnitude  
between our two simulated galaxies, from $\sim 50$\% in G1 to
$\sim 5$\% in G2. 
Note that our simulations do not include ionising radiation from the stellar
sources (or from a UV background). This missing ingredient could
increase the escape fractions, and possibly reduce or the difference
between G1 and G2. We plan to address this issue in a forthcoming
paper.

\vskip 0.2cm

Fig. \ref{fesc_jeje} also shows color histograms in each panel. These  
represent the escape fraction distributions for different populations  
of stars. The red (resp. yellow, green) curves show the escape  
fractions for photons emitted in high (resp. medium, low) density  
environments\footnote{The cuts in local stellar density that we used 
vary from G1 to G2, and are defined arbitrarily to capture the different 
regimes illustrated in Fig. \ref{fesc_jeje}.}. 
Thumbnail images in the lower left corner of each  
panel show where the sources contributing to each curve are located in  
the galaxies (see Figs. \ref{wind} and \ref{clumps} for comparison).  
The very strong variation of escape fractions with environment in G2,  
as opposed to the very small variation in G1, is the key to the  
difference between these two galaxies, and explains our disagreement  
with the results of \citet{Laursen2009}. This different behavior was  
expected, since the ISM in G1 (along with most simulations in the  
literature) is artificially smoothed
on small scales by the unrealistically large pressure support  
associated with 10$^4$~K gas. Hence, the weak radial variation in the  
escape fraction of G1 simply reflects the gas density gradient in that  
direction. In G2 however, most young stars lie within very dense and  
dusty clouds, as expected from e.g. \citet{CharlotFall00}, and  
consequently their \lya{} radiation is  heavily absorbed. These  
results demonstrate the necessity to resolve (at least some of) the  
ISM structure of galaxies with hydrodynamical simulations before post- 
processing them with \lya{} radiative transfer codes in order to make  
more realistic predictions for the \lya{} escape fraction.

It may seem surprising that our results show the opposite trend as  
that expected in the scenario advocated by \citet{Neufeld91}: we find  
that a clumpy ISM reduces the escape fraction of \lya{} photons  
relatively to continuum photons instead of enhancing it. The reason  
for this behavior is that our G2 simulation predicts a configuration  
which is quite different from that assumed by \citet{Neufeld91}.  
Whereas this author studies the propagation of photons {\it through} a  
clumpy medium, we find that our photons are actually mostly {\it  
emitted within the clumps}, and it is their (in)ability to escape from  
these clumps that dominates the results. Clearly, our G2 simulation  
still suffers from limitations, both in terms of the simplified  
assumptions we make to describe the ISM physics (no UV RT but CIE)
and in terms of the resolution, which is insufficient to fully resolve 
\hii{} regions.  
Bearing these caveats in mind, we find no support for the `Neufeld  
scenario'. We plan to return to this issue with improved simulations  
in future work.

\subsection{\lya{} spectra}
\label{s_spec}
\begin{figure}
\includegraphics[width=8.6cm]{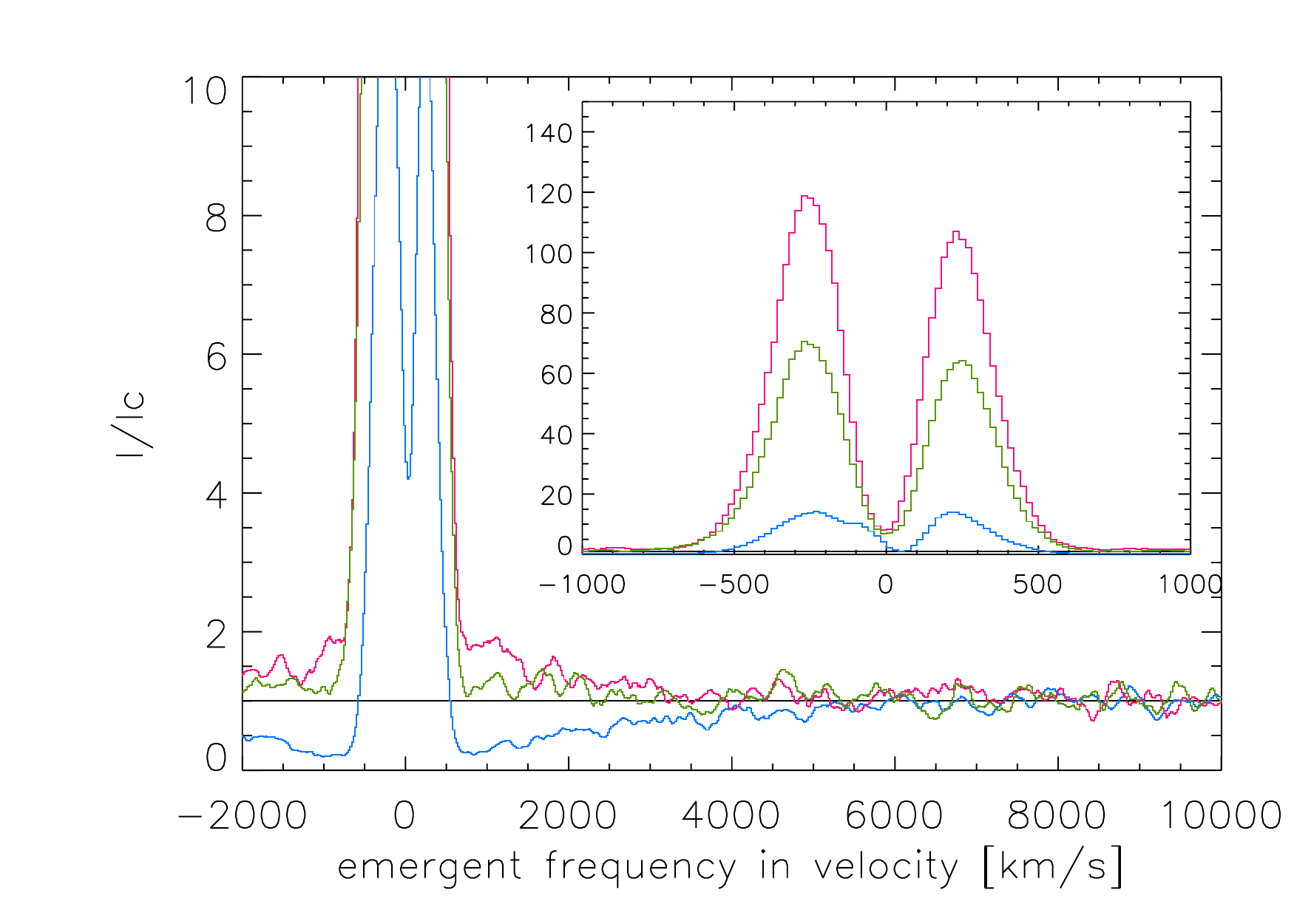} \\
\includegraphics[width=8.6cm]{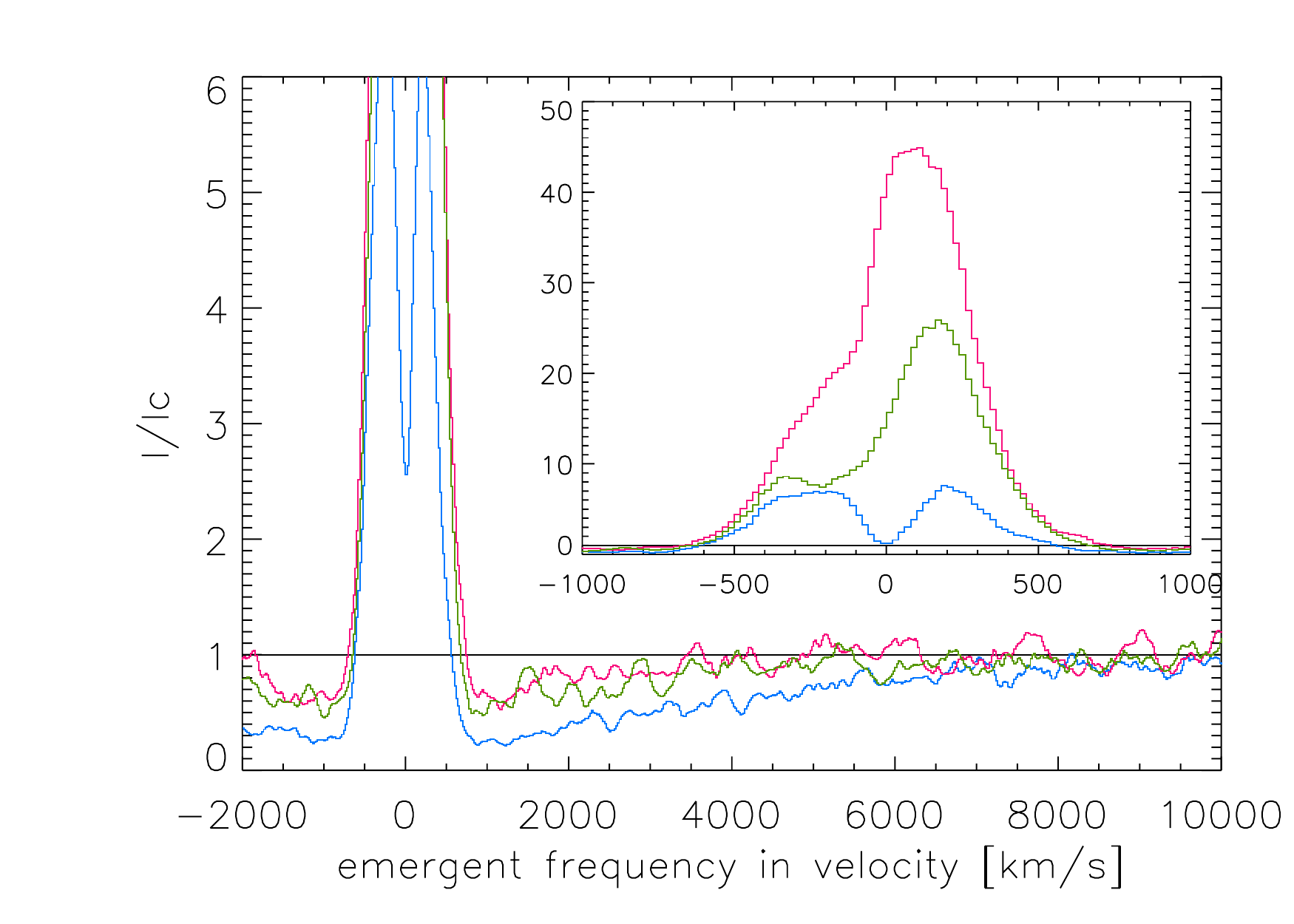}
\caption{{\bf Top :} Emergent spectra from G1 along 3 lines of sight 
($\theta=0$ in magenta, $\theta=\pi/4$ in green, $\theta=\pi/2$ in blue).
All spectra are \emph{spatially integrated}, 
i.e. they are emergent spectra along a given direction, 
but from the whole galaxy. 
{\bf Bottom :} Same as top panel but for G2.}
\label{spectra_line}
\end{figure}

We now turn to our predicted spectra, i.e. the distribution of {\it  
observer-frame} frequencies of photons which escape G1 and G2. We show  
these spectra in Fig. \ref{spectra_line} for three different  
inclinations: face-on (magenta curves), edge-on (blue curves), and seen  
from 45 degrees (green curves).
Thanks to the symmetry of the problem, we checked that 
the emergent spectra in all azimuthal directions are the same. 
The emergent spectra escaping at inclinations $\pm cos\theta$ are also identical. 
To increase the statistics, we then sum all photons escaping with $|cos\theta|$
in the bin corresponding to the chosen inclination, whatever its azimuthal angle $\phi$ is
The spectra presented on Fig~\ref{spectra_line} are built by
collecting photons within $cos\theta$ bins large enough to a robust
signal, but small enough to keep all the angular variations
(e.g. $|cos\theta| > 0.95$ face-on and $|cos\theta| < 0.05$ edge-on).
Our spectra are \emph{spatially integrated}, 
i.e. they are emergent spectra along a given 
direction but from the whole galaxy.

\begin{figure}
\begin{center}
\includegraphics[width=8.6cm]{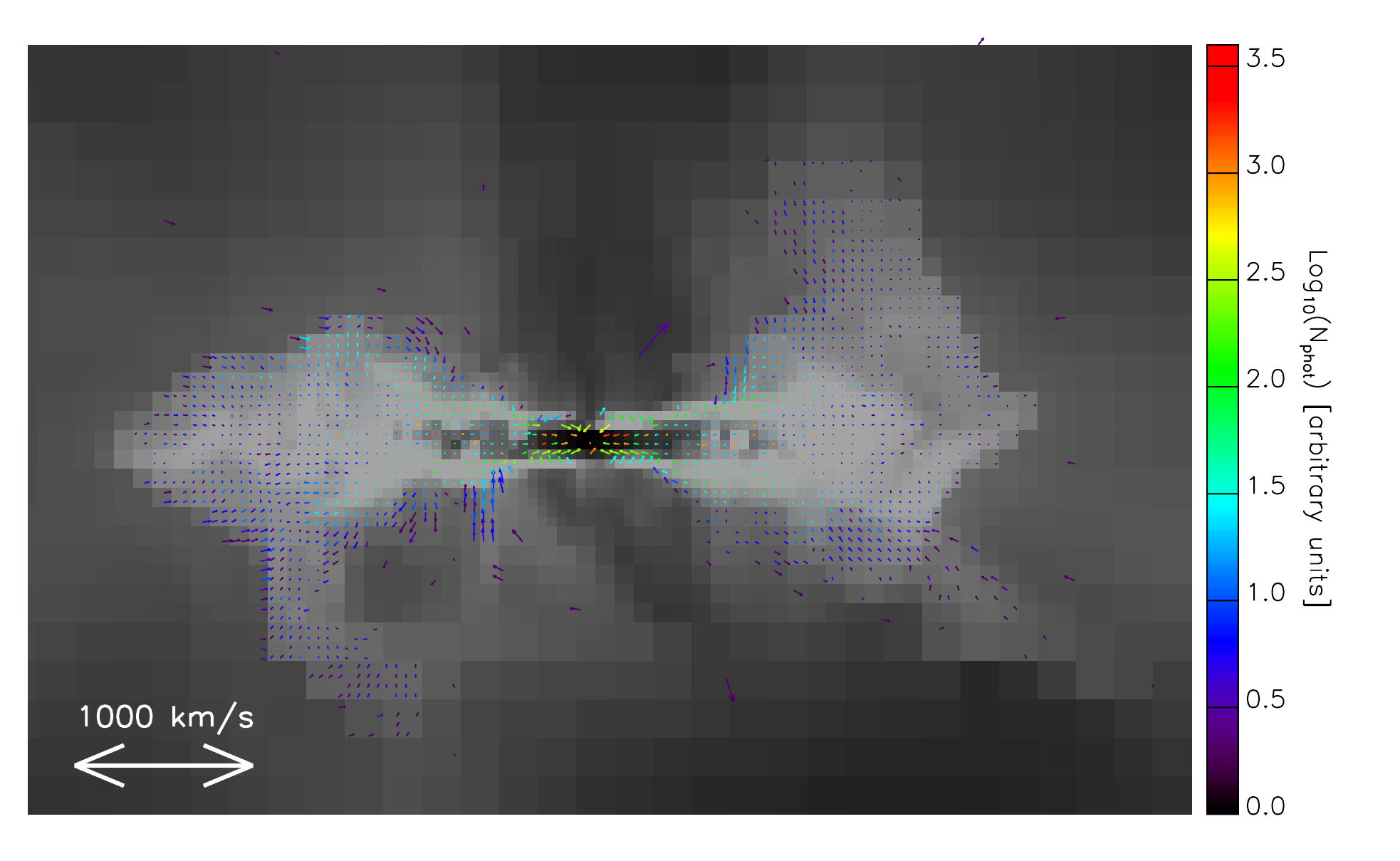} \\
\includegraphics[width=8.6cm]{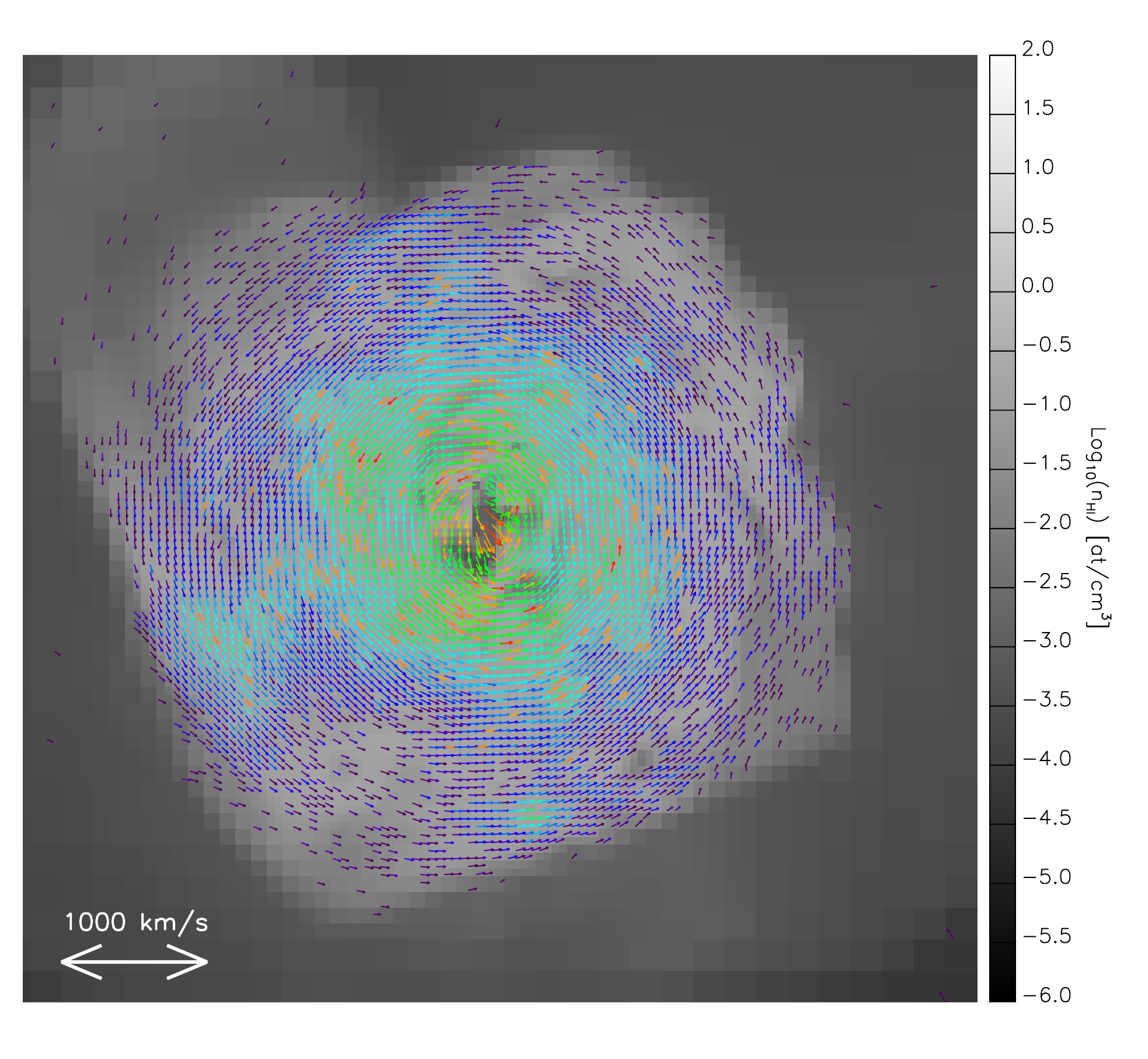}
\end{center}
\caption{Edge-on (top) and face-on (bottom) views of G1. In both  
panels, the grey scale indicates the density of neutral Hydrogen  
(maximum value in a slice of thickness 8 high-resolution cells). The  
arrows indicate the velocity field of the gas at the location where  
photons last scatter before they are observed (the double sided arrow  
in the bottom left corner of each panel gives the scale), and their color 
scales with the log$_{10}$ of the number of photons seen from each cell 
(red: lots of photons, blue: small number of photons). }
\label{fig:velocity_field_g1}
\end{figure}

Looking at the spectra emerging from G1 in the top panel of Fig.  
\ref{spectra_line} we see that they are all double-peaked, and roughly  
symmetrical about the line center. The overall intensity (and the  
equivalent width) decreases regularly from the face-on to the edge-on  
view. This is expected since the typical optical depth from the  
sources to the observer varies in the same way. The double-peaked  
symmetrical line resembles that expected from transfer through a  
static homogeneous medium. This is not surprising, since the ISM  
of G1 is rather homogeneous with a velocity field typical of a  
globally rotating disk with little internal motions 
(cf Fig~\ref{fig:velocity_field_g1}, bottom panel). The somewhat  
higher intensity in the blue peak (at negative velocities) is the  
result of transfer effects through the gaseous halo, which is in part  
in-falling towards the galaxy (cf Fig~\ref{fig:velocity_field_g1}, top panel).
Indeed, we checked that the spectrum of photons emerging 
face-on and for which the last scattering is in the halo 
(above the disk plane) 
is more asymmetric than that of photons emerging directly from the disk.

\begin{figure}
\begin{center}
\includegraphics[width=8.6cm]{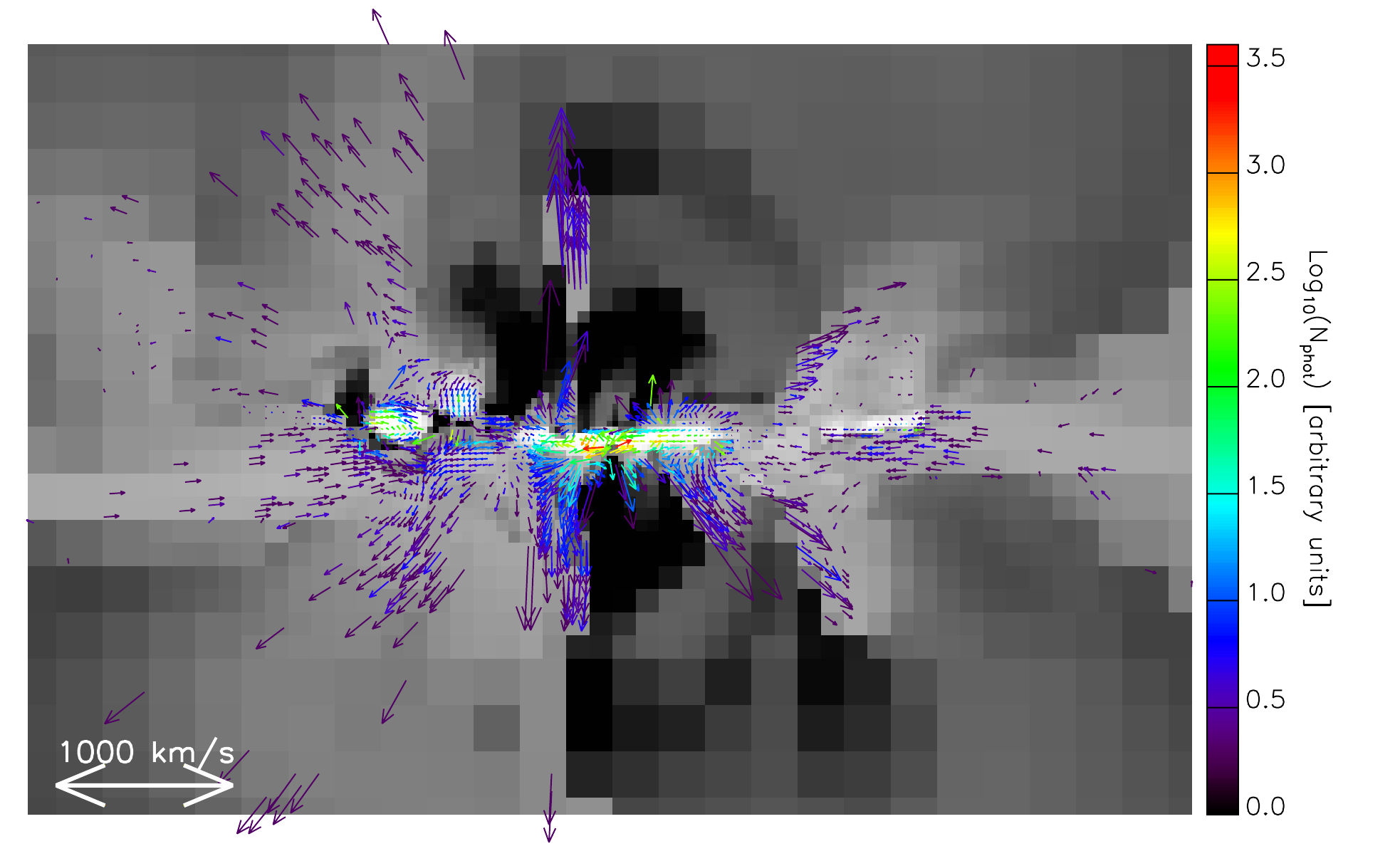} \\
\includegraphics[width=8.6cm]{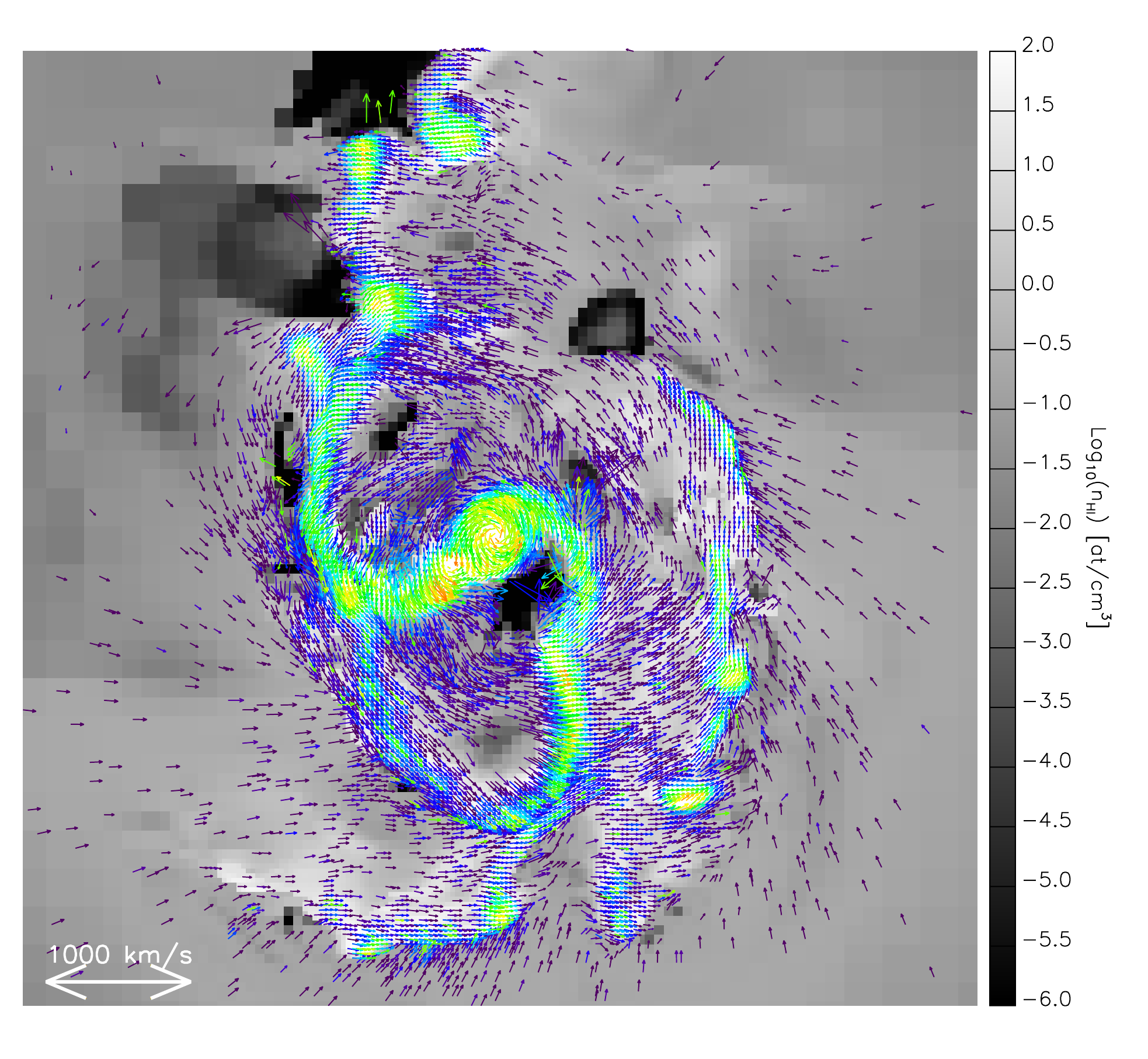}
\end{center}
\caption{Same as Fig. \ref{fig:velocity_field_g1}, for G2. Note the  
much more pronounced vertical motions in the top panel.}
\label{fig:velocity_field_g2}
\end{figure}

The emergent spectrum from G2 (bottom panel of Fig.  
\ref{spectra_line}) is also double-peaked edge-on, but clearly  
asymmetric face-on.  The enhancement of the red peak compared to the  
blue peak is a signature of \lya\ diffusing in an outflowing medium, 
and indeed, gas in G2 is outflowing at much larger
  velocities (of order a few 100 km/s) than in G1 (where velocities
  are close to zero or infalling, see Fig~\ref{fig:velocity_field_g1} and
Fig~\ref{fig:velocity_field_g2}, top panels).   
We checked that if we set the velocity field to 0 \kms{} in each cell  
of the simulation, the spectrum of G2 seen in any direction becomes  
symmetric about the line center ($v=0$ \kms). We also reversed the  
vertical component of the velocity field, $v_z$, and checked that in  
this case, the emergent spectrum is reversed, with a more prominent  
blue peak. Although a very strong large-scale galactic wind is present  
in G2, this outflowing material is mostly ionized and tenuous,  
conditions which render it transparent to \lya{} photons. As can be  
seen from Fig.~\ref{clumps}  (bottom left panel), SN feedback also  
pushes cold neutral gas outside the disc, albeit not very far.  
However, even such a small-scale kinematic feature is enough to  
produce the asymmetry observed in Fig. \ref{spectra_line}, 
because the kinematic energy transfered to this neutral gas 
is significant (cf Fig~\ref{fig:velocity_field_g2}, top panel).

Even if the asymmetry of the G2 spectra qualitatively compares well to  
observations \citep[e.g.][Fig 10]{Kulas2012}, the shift of the peak and 
the extension of the red tail  
are less important than in most observed spectra \citep[e.g.][Fig 1] 
{Steidel2010}.  We attribute (at least part of) this discrepancy  to  
the mass of the galaxy we simulated. This latter is about two orders of  
magnitude less than typical Lyman-break galaxies, and therefore we can  
hardly expect its kinematics to have the same amplitude. 
A wrong implementation of the wind mechanism could 
cause the same mismatch. The simplified assumptions used to describe 
the ionisation state of the interstellar gas, i.e. CIE, can also 
lead to an overestimate of the neutral fraction of the interstellar medium, 
especially in G2, and affect the spectral shapes. 
However, investingating different feedback recipes and following UV ionising 
radiation transfer is beyond the scope of this paper.

\section{Orientation effects}

We discussed above the impact of the small-scale ISM structure on the 
propagation and escape of \lya{} photons. In the present section, we inspect 
the global effect of inclination on the observed \lya{} properties of our 
simulated galaxy G2. Indeed, orientation effects were predicted by
\citet{CharlotFall93, ChenNeufeld94} and pointed out by recent 
studies\citep{Laursen2007,Laursen2009,Zheng2010,Barnes2012,Yajima2012}.

\subsection{Angular escape fraction and angular redistribution}

Let us start by noting that the observed \lya{} properties of G2 are,
as expected, invariant around the axis of rotation of the disc,
i.e. they do not depend on the azimuthal angle. They do however
strongly depend on the inclination angle $\theta$, which we define
here as the angle from the rotation axis of the disc ($\theta=0$
face-on and $\theta = \pi/2$ edge-on).

\begin{figure}
\includegraphics[width=8.6cm]{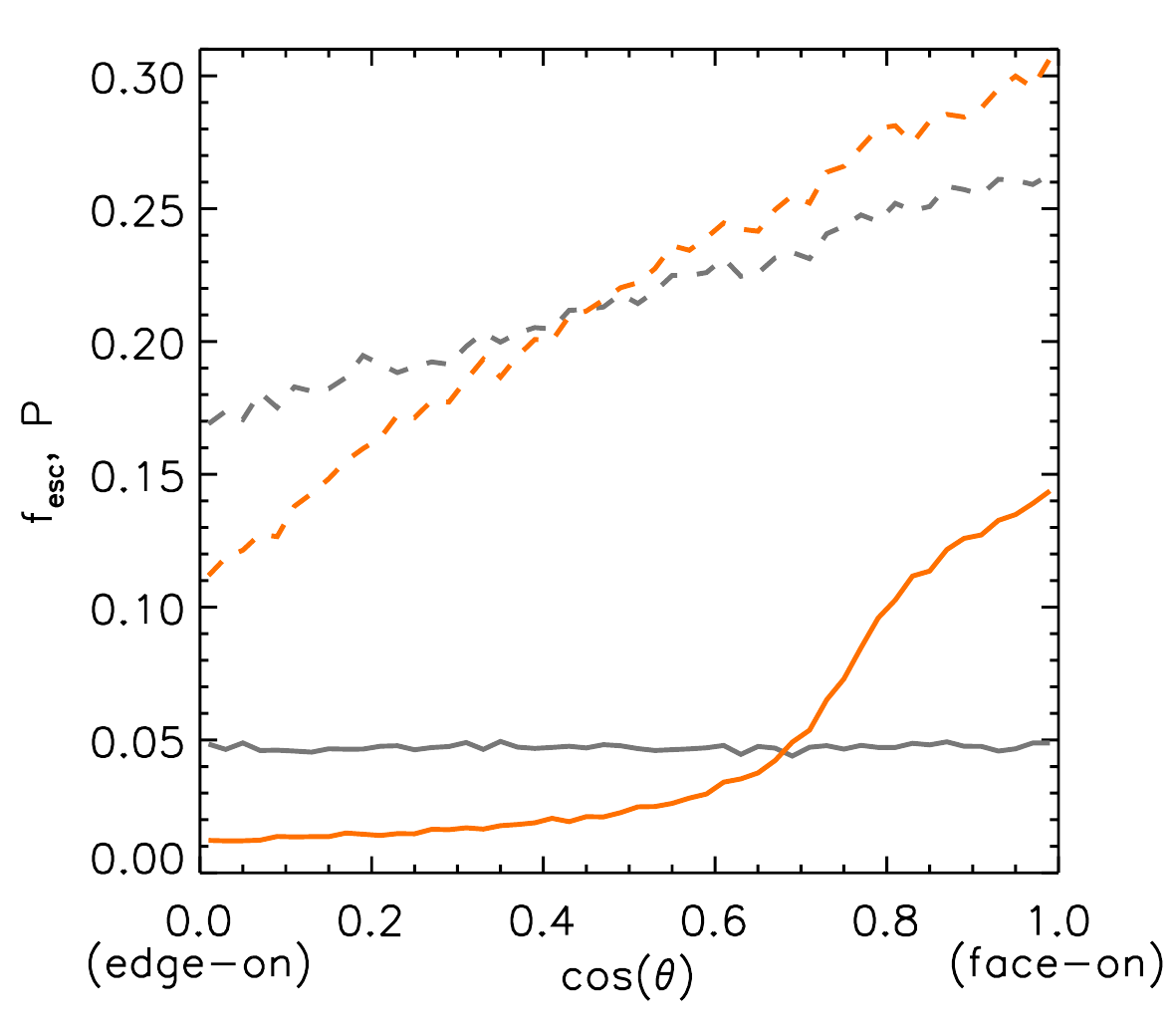}
\caption{Comparison between the distributions of escape directions in orange, 
and emission directions in grey, for line (solid lines) and continuum 
(dashed lines) escaping photons. See the text for a detailed explanation of 
the shapes of these distributions.
Note that the y axis has two different meanings, $f_{esc}$ for 
the distribution of emission directions, and P for the normalised distribution of escape directions.}
\label{angles}
\end{figure}

The grey curves in Fig. \ref{angles} show the angular escape fractions
of continuum (dashed line) and line (solid line) photons\footnote{We
  define line photons as having $|v_{\rm em}| < 100$~km/s in the frame
  of the sources (see black curve of Fig. \ref{fig:input_spec}). For
  continuum photons, we take photons with $|v_{\rm em}| > 7000$~km/s.}. 
These escape fractions are
computed as a function of the inclination at which photons are
emitted. The flat solid grey line thus tells us that the escape
fraction of line photons is isotropic: whatever direction they are
emitted along, only about 5\% will escape. On the contrary, continuum
photons tend to escape more easily when emitted perpendicular to the
disc. Their escape fraction varies from 17\% when emitted in the plane
to 26\% when emitted face-on. The difference between these behaviors
is due to resonant scattering.  Most line photons scatter a huge
number of times (70\% scatter more than $10^6$ times, less than 10\%
scatter less than 10 times), enough to forget their initial direction.
On the contrary, continuum photons which escape do so without
scattering (about 65\% escape directly, and all escape with less than
8 scatterings).

The orange curves in Fig. \ref{angles} show the distributions of the
inclinations with which the photons escape, i.e. the observed
inclination angle distribution, again for continuum (dashed line) and
line (solid line) photons. The solid orange curve of Fig. \ref{angles}
shows a strong angular redistribution of line photons: although they
are emitted isotropically and have an isotropic escape fraction, the
probability (per unit solid angle) of one escaping face-on is about 15
times higher than that of one escaping edge-on. Similarly, albeit with
a much lower amplitude, continuum photons are also redistributed in
direction, so that they have a probability 3 times higher to come out
face-on than edge-on. 

\begin{figure}
\includegraphics[width=8.6cm]{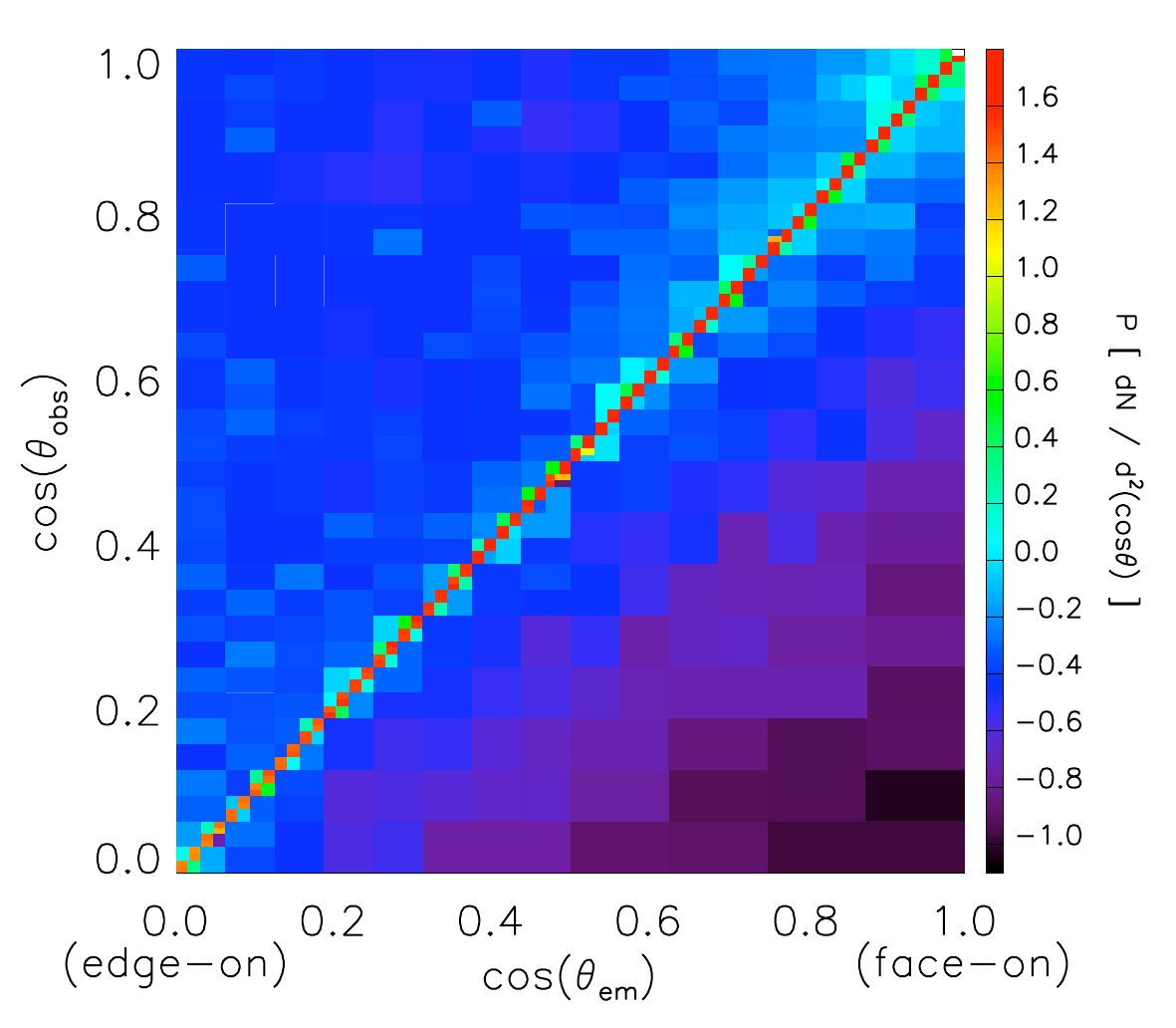} 
\includegraphics[width=8.6cm]{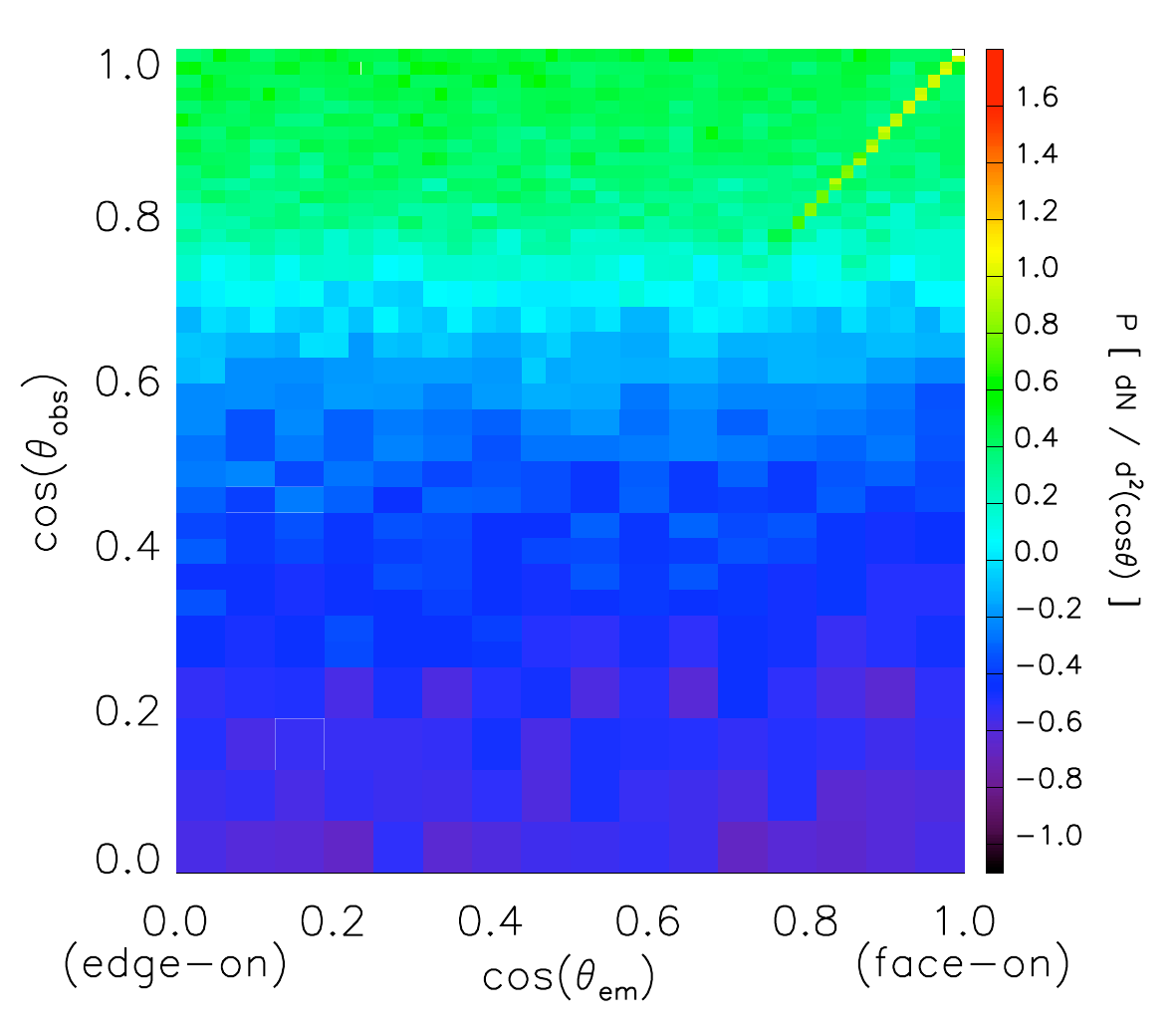}
\caption{Angular redistribution of continuum (top panel) and line 
(bottom panel) photons. The x-axis gives the emission direction 
(as the cosine of the inclination angle), and the y-axis gives the 
direction along which the photons escape. The color coding scales 
logarithmically with the number of photon per unit area in the plot, 
as indicated by the color tables. The white lines on both panels, 
show the escape fraction as a function of the emission direction. }
\label{fig:ang_redist}
\end{figure}

To understand better this angular redistribution, we show in
Fig. \ref{fig:ang_redist} the relation between observed (y-axis) and
emitted (x-axis) inclinations of the photons which escape. The top
panel of Fig. \ref{fig:ang_redist} shows that indeed, most continuum
photons are observed with the same inclination as they were emitted
with. This panel also shows that a small fraction of continuum photons emitted edge-on
are scattered and observed as coming from any direction. The lower
panel of Fig. \ref{fig:ang_redist} shows that the situation is quite
different for line photons. Here, photons only have a marginal chance
to escape in the same direction as they were emitted in, provided they
were emitted almost face-on, and the overall redistribution is almost
independent on the emission inclination. This is due, of course, to
the very large number of scatterings that line photons undergo in
general. 

An early prediction of such orientation effect 
(favoring face-on escape) was made by \citet{CharlotFall93}, 
by analogy with the stellar 'limb-darkening' effect.
However, the enhancement factor that we get ($\sim15$) 
is much higher than their prediction ($\sim2.3$).

\vskip 0.2cm This idealised simulation of galaxy formation
may be a much more symmetrical configuration than galaxy formation in a 
cosmological context. However, if still relevant for real galaxies,
the strong orientation effect we find has important
observational implications. In particular, our results suggest that an
LAE survey, which introduces a \lya{} luminosity selection, will be biased
towards face-on objects, and much more so than e.g. Lyman break
surveys which rely on a (UV continuum) broad-band selection. The LAE
may hence represent only an incomplete survey of star
forming objects at high redshifts. Also, we
expect inclination to introduce a significant scatter in correlations
between e.g. SFR and observed \lya{} luminosity. 

\subsection{\lya\ equivalent widths}
\label{s_EW}

\begin{figure}
\includegraphics[width=8.6cm]{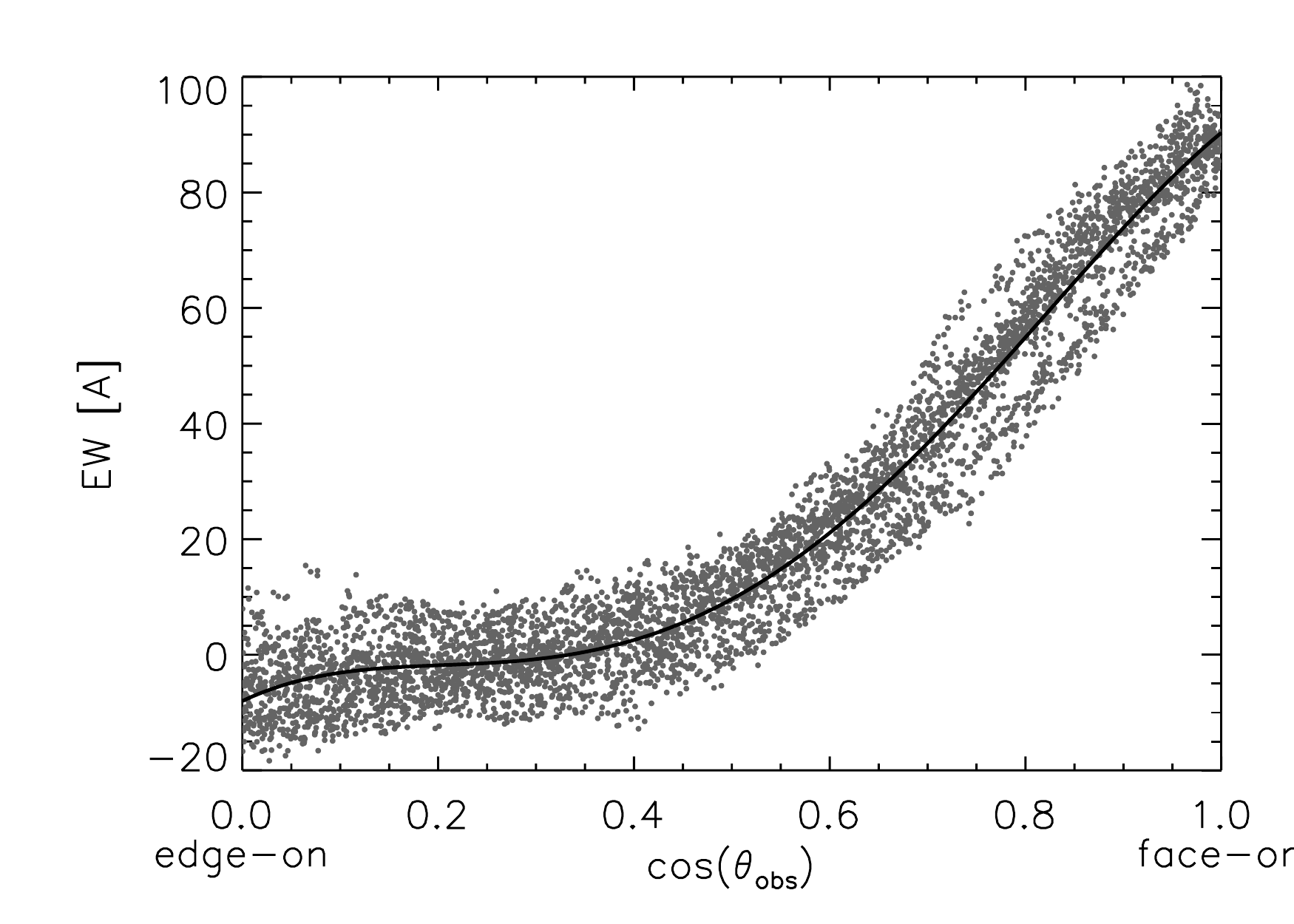} 
\caption{\lya\ equivalent widths as a function of the (cosine of the)
  inclination at which our simulated galaxy is seen. Each point marks
  one of 5000 mock observations. The solid line is a simple polynomial
  fit (see text).}
\label{inclination}
\end{figure}

The inclination effect discussed above also manifests itself in the
form of a strong correlation between the observed equivalent width and
the inclination at which the galaxy is observed. In
Fig. \ref{inclination}, we show the EW as a function of the
inclination angle, each point corresponding to a random direction of
observation. The tight correlation is well fit by a polynomial of the
form $EW(\mu) = -8 + 80.7\ \mu - 393.1\ \mu^2 + 798.2\ \mu^3 - 387.4\
\mu^4$, where $\mu = \cos(\theta)$. This fit is shown with the solid
black line. The scatter across the relation is due to variations
in the azimuthal angle ($\phi$). Note that the EWs we compute here
only take into account continuum radiation from stars younger than
10~Myr (the same that are used as sources for \lya{} photons), which
is good approximation. 

The strong dependency of the EW on $\mu$ illustrates the differential
effect introduced by resonant scattering and discussed with
Fig. \ref{angles}. It shows the complexity of the possible bias
introduced by EW selections in narrow-band\footnote{Although narrow
  band surveys introduce by construction an EW selection, most
  spectroscopic searches bear such a bias as well.} LAE samples: not
only does the \lya{} luminosity of our simulated galaxy strongly vary
with inclination, the EW of its emission line does too, and as
strongly. For example, with a typical selection at EW~$>20$~\AA{}, our
simulated galaxy would be found as a LAE only if observed with an
inclination $|\mu| \gtrsim 0.6$, i.e. with $\sim 40$\% chance assuming
a random inclination. In other words, if real galaxies behave
similarly to our simulation, LAE narrow-band surveys could be missing
$\sim 60$\% of the faint galaxy population at high redshifts. 
Even if this number is probably an upper limit, given the symetry of our 
idealized galaxy , it calls for observational studies which could assess 
the effect of inclination on the \lya{} properties of real galaxies.

\subsection{Line profile}

As presented on Fig~\ref{spectra_line}, the emergent spectrum from our
realistic ISM model G2 shows a transition in its shape with
inclination: the edge-on profile is a double peak symmetrical around
the line center, and the face-on spectrum has a much attenuated blue
peak. The origin of this behavior is mainly illustrated in
Fig. \ref{fig:velocity_field_g2} and is as follows. \lya{} photons seen
escaping edge-on have scattered through an extremely thick medium,
with little radial velocities, and the line they produce thus
resembles the double peak of a static medium in which photons scatter
until they are shifted far enough from the line center to
escape. \lya{} photons seen face-on, on the contrary, traverse a much
lower column density with a much more pronounced velocity field, with
gas outflowing at velocities of several 100 km/s, which favors the
escape of red-shifted photons. 

This evolution of the line profile suggests that the integrated
spectral shapes of galaxies may reflect to some extent their
inclination, as much as their internal structure and
kinematics. Clearly, one would need a larger sample of simulated
galaxies to draw robust conclusions here. However, this trend found in
our results does hint towards yet another selection effect: Our
simulation predicts a correlation between the EW and the line profile,
due to the strong correlation that both these quantities have with
inclination. This implies that an EW selection will favor asymmetric
line profiles and tend to exclude double peaks, which we find to be
associated to low inclinations.

\section{\lya{} halo}

\begin{figure}
\includegraphics[width=8.6cm]{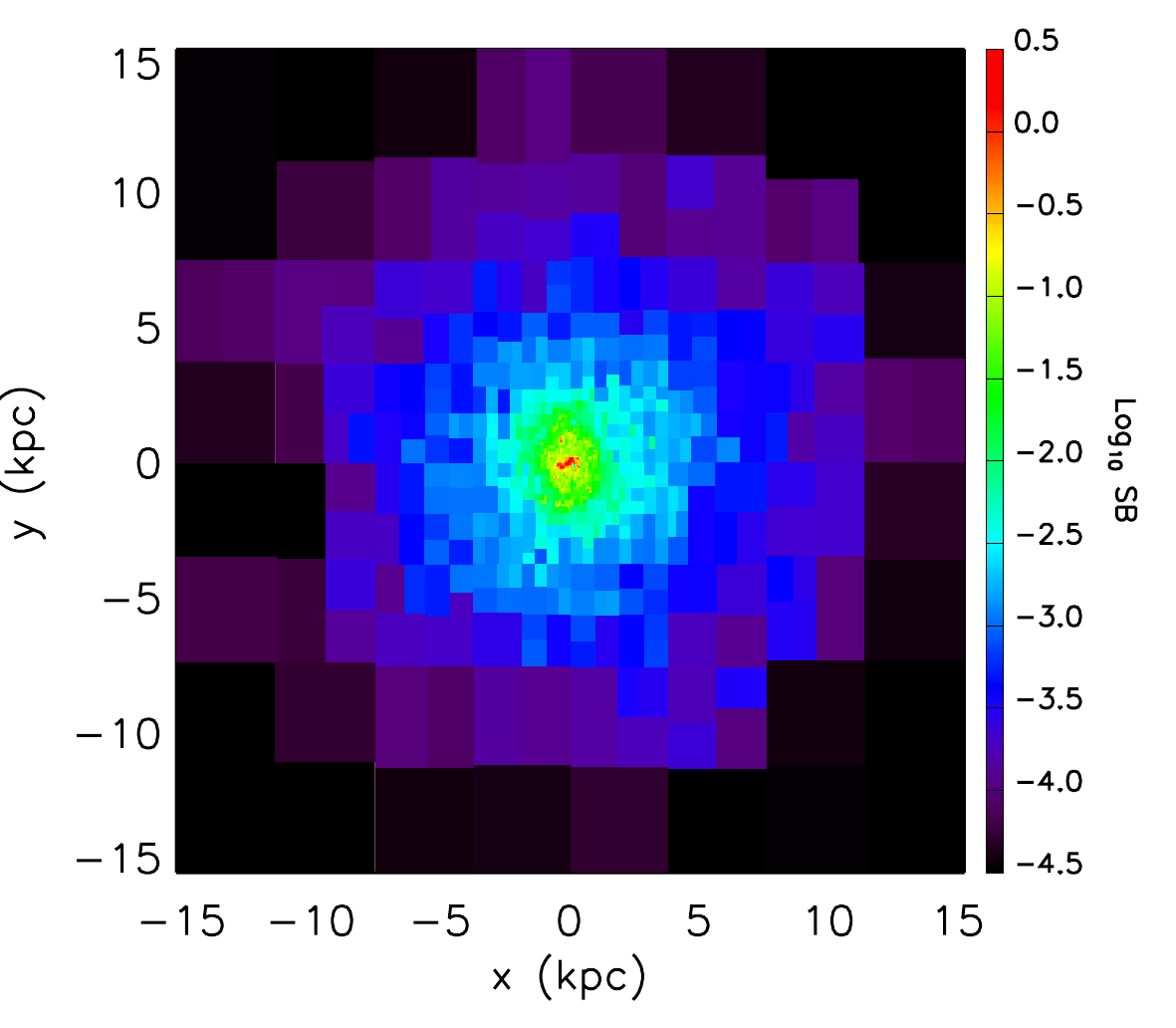} \\
\includegraphics[width=8.6cm]{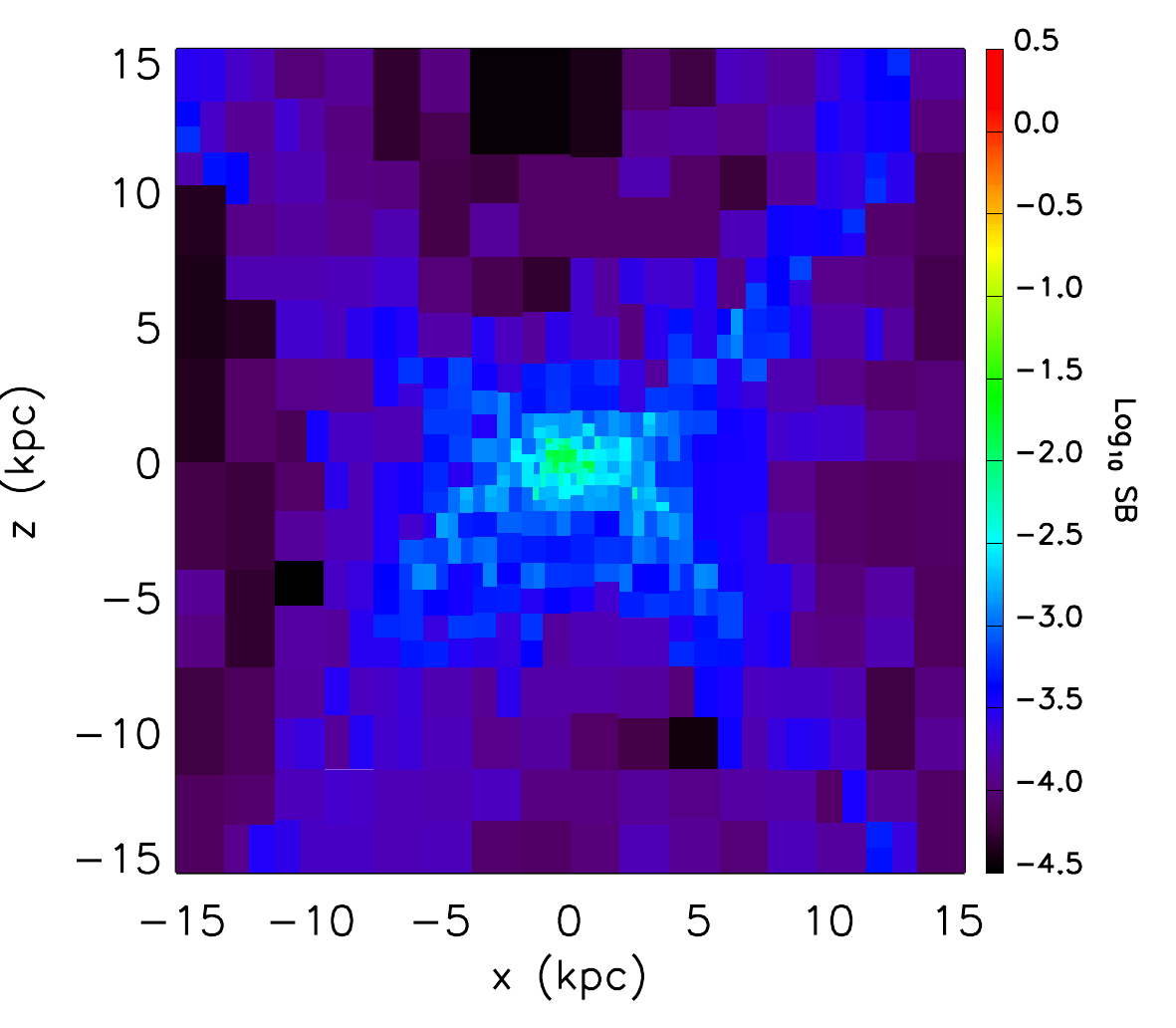}
\caption{Face-on (top) and edge-on (bottom) views of simulated galaxy
  G2 in \lya{}. The color-coding shows surface brightness in arbitrary
units, with the same scale in the two panels.}
\label{im_G2}
\end{figure}

On Fig~\ref{im_G2}, we present face-on (top panel) and edge-on
(bottom panel) images\footnote{Images along a given 
line of sight described by $\vec k_{los}$ are obtained 
by selecting photons escaping in a cone of angle $\alpha$ defined by 
$ -1 < \vec k_{los} . \vec k_{out} < cos(\pi-\alpha)$, 
with $\alpha$ being as small as possible to achieve the best accuracy, 
and high enough to collect a significant number of photons.} 
of the galaxy G2 in the \lya\ line.  
Scattering of \lya{} photon through the tenuous intra-halo medium produces a
diffuse \lya{} halo which is clearly visible from both angles. The face-on
view only includes photons with $|\cos(\theta)| > 0.95$. The
edge-on view includes all photons with escaping
$|\cos(\theta)| < 0.2$ so as to have enough statistics, and its
surface brightness was renormalized accordingly, so that both panels
in Fig.\ref{im_G2} are directly comparable. 
The observed \lya\ luminosity face-on 
is 15\% of the intrinsic luminosity (cf Fig~\ref{angles}, 
$L(\lya)_{\rm obs} = 0.15\times L(\lya)_{\rm int} \sim 1.8\times10^{41}$ erg.s$^{-1}$), 
whereas it is only a few percent in the edge-on sightline
($\sim 1-2\times10^{40}$ erg.s$^{-1}$). 
Such faint galaxy can be observed from $z=0$ 
\citep[][Lyman Alpha Reference Sample,]{Ostlin2009}, 
until $z\sim3$ \citep{Rauch2008,Garel2012}.
As can be seen from the edge-on view, most of the extended emission comes 
from scattering on the envelope of the galactic wind produced by G2, 
and the hot wind itself is mostly transparent. 

This picture is qualitatively comparable to observations by e.g.
\citet{HayesEtal05,Ostlin2009} or \citet{Steidel2011}. Quantitatively, we find
that about 40\% of \lya{} photons come from the central region (within
1.5~kpc), and the rest is diffuse. This does not fit well with the
results of \citet{Steidel2011}, who find about 5 times more luminosity
in the extended halo than in the central region. There are however a
number of likely explanations to this apparent disagreement. First,
our simulated galaxy is about two orders of magnitude less massive than
a typical LBG. Second, and probably more important, our simulation is
idealized, and the circum-galactic medium does not feature cold
streams or tidal tails which would enhance the effect of
scattering. Third, our calculations do not include in-situ emission
from cold gas in the halo, either from cooling radiation 
or UV background fluorescence, and this may play a dominant role here
\citep[e.g.][]{RosdahlBlaizot2012}.

\section{Summary and conclusion}
\label{s_conclude}

In this paper, we have studied the \lya{} properties of a couple of
high-resolution simulated dwarf galaxies forming in an idealized dark
matter halo.  Our two simulations assume different temperature floors
of the cooling function (10$^4$K for G1, 100K for G2), which result in
strikingly different structurations of the ISM. While the gas in G2 is
able to fragment into small star-forming clumps, the thermal pressure
support in G1 yields a rather smooth ISM with homogeneous star
formation.  We have post-processed these galaxies with \McLya{} in
order to follow the resonant scattering of \lya{} photons through
their ISM, and to predict their resultant observational
properties. Our main results are as follows.
\begin{itemize}
\item[1.] As expected, the small scale structure of the ISM plays a
  determinant role in shaping a galaxy's \lya{} properties. In the G2
  simulation, where gas is allowed to cool down to temperatures $\ll
  10^4$ K, most young stars are embedded in thick, dusty, star forming
  clouds, and the photons they emit are strongly attenuated. As
  opposed to the ``Neufeld scenario'', the {\it clumpiness of the ISM
    here enhances the destruction of \lya{} photons relative to
    continuum photons}. This is due to the fact that photons are
  emitted within the dense clouds, {\it \`a la} \citet{CharlotFall00},
  rather than outside, as assumed in \citet{Neufeld91}.

  In the G1 simulation, with an artificially warm ISM, young stars are
  found in much lower density environments and their photons escape
  more easily. This simulation is comparable to the previous studies
  in the literature which also include dust
  \citep{Laursen2009,Yajima2011,Yajima2012}, and our results are
  indeed similar to these studies. Such simulations do not capture the
  enhancement of \lya{} extinction relative to continuum in
  star-forming regions that we find in G2, simply because they do not
  form such dense star forming regions.

  Another important feature is the kinematic structure of the
  ISM. Because G2 develops a genuine multiphase medium, with very
  dense star forming clouds and a tenuous diffuse component,
  supernovae explosions are able to push gas to high velocities (see
  Fig. \ref{fig:velocity_field_g2}). Instead, the rather homogeneous
  ISM of G1 is overall denser than the diffuse medium of G2, and
  resists better supernovae explosions. It thus displays a rather
  static velocity field (see Fig. \ref{fig:velocity_field_g1}). As
  shown in Sec. \ref{s_spec}, these different velocity fields have a
  strong impact on the shape of \lya{} lines.

\item[2.] The analysis of \lya{} emission from G2\footnote{and G1,
    although we do not show it here} demonstrates the existence of
  {\it a strong inclination effect}. Due to the numerous scatterings
  that line photons undergo, their probability to escape does not
  depend on the direction towards which they were emitted. Instead,
  they tend to systematically escape face-on, following the path of
  least opacity. Because continuum photons do not display such a
  strong angular redistribution, this effect is directly seen on the
  \lya{} equivalent width, which we find to vary from $\sim -5$ \AA{}
  edge-on to $\sim 90$ \AA{} face-on. 
  
  We also find that this inclination effect is seen in the shape of
  the \lya{} line emerging from our simulated galaxy. When seen
  edge-on, our galaxy has a double-peak line, associated with a low
  EW. When seen face-on, our galaxy has an enhanced red peak, and a
  high EW. 

  These results suggest the possible existence of strong observational
  biases in LAE surveys which necessarily rely on \lya{} luminosity
  and EW selections, and could thus preferentially select face-on
  objects. As an example, a survey with an EW cut at $~20$~\AA{} would
  select our galaxy only 40\% of the times, assuming it has a random
  inclination.

\item[3.] Scattering of galactic \lya{} photons through the
  circum-galactic medium do produce an extended \lya{} halo. We find
  that about a third of \lya{} photons escaping G2 contribute to this
  diffuse component. This is somewhat at odds with the results of
  \citet{Steidel2011}, though the comparison should be taken with
  caution given the fact that our simulated galaxy has a much smaller
  mass than the galaxies analyzed by these authors. Also, our
  simulation is idealized, and the CGM of G2 is not a good
  representation of what one finds in cosmological simulations at high redshift
  \citep{RosdahlBlaizot2012, Dubois12}.
\end{itemize}

Although our quantitative results are clearly limited by much missing physics,
we believe that this work demonstrates the two main following points: 
(1) resolving the ISM is mandatory if we want to understand the escape 
fraction of Lya from galaxies (it doesn’t matter here if we have the correct 
solution: we show how widely the results vary when we change the ISM 
physics ... and this definitely shows that we should go further); 
and (2) for an ideal disc galaxy, we find that the escape fraction is a 
strong function of inclination, and we argue that this effect is quite 
possibly present in real galaxies (eventhough their morphologies are known to
be more complex). Both results call for more work, both
theoretically and observationally. From the theoretical viewpoint,
we plan to make progress in forthcoming papers by (i) including the
transfer of ionizing photons through a {\it resolved} ISM, and (ii)
embedding our galaxy in the full complexity of its cosmological
context.

\acknowledgements{
  AV was supported by a Fellowship for prospective researchers of the
  Swiss National Science Foundation to start this project in Oxford,
  and by a Marie Curie Intra European Fellowship within the 7th
  European Community Framework Programme in Lyon. YD is supported by
  an STFC Postdoctoral Fellowship.  The simulations presented here
  were run on the TITANE cluster at the Centre de Calcul Recherche et
  Technologie of CEA Saclay on allocated resources from the GENCI
  grant c2009046197. JB acknowledges support from the ANR BINGO
  project (ANR-08-BLAN-0316-01).

}

\bibliographystyle{plainnat}
\bibliography{references}

\end{document}